\documentclass[conference]{IEEEtran}
\usepackage[utf8]{inputenc}
\usepackage{color}
\usepackage{float}
\usepackage{tabularx}
\usepackage{graphicx}
\usepackage{adjustbox}
\usepackage{graphicx}
\usepackage{subcaption}
\usepackage{comment}
\usepackage{hyperref}
\usepackage{amsmath}
\usepackage{amsfonts}
\usepackage{tikz}
\usepackage{authblk}
\def\checkmark{\tikz\fill[scale=0.3](0,.35) -- (.25,0) -- (1,.7) -- (.25,.15) -- cycle;}

\begin{document}
\IEEEoverridecommandlockouts

\title{The backpropagation-based recollection hypothesis: \\[0.2em]
\LARGE Backpropagated action potentials mediate recall, \\ 
			imagination, language understanding and naming}		       

\author{Zied Ben Houidi$^{1}$\thanks{$^{1}$zied.benhouidi@gmail.com, zied.ben.houidi@huawei.com}}
\affil{\textit{Huawei Technologies Co. Ltd.}}

\maketitle
\begin{abstract}
Ever since the advent of the neuron doctrine more than a century ago, information processing in the brain is widely believed to mainly follow the forward pre to post-synaptic neurons direction. 
Challenging this prevalent view, in this paper,
we emit the \textit{backpropagation-based recollection hypothesis} as follows: weak and fast fading Action Potentials following the (highest weight) post to pre-synaptic backward pathways, mediate explicit cue-based memory recall. This includes also the tasks of imagination, future episodic thinking, language understanding and associating names to various stimuli. 
These signals originate in highly invariant neurons, which uniquely respond to some specific stimuli (e.g. image of a cat). They then travel backwards to reactivate the same populations of neurons that uniquely respond to this specific stimuli during perception, thus recreating ``offline'' an experience that is similar.  
After stating our hypothesis in details, we challenge its assumptions through a thorough literature review. We find abundant evidence that supports most of its assumptions, including the existence of backpropagating signals that have interesting properties. 
We then leverage simulations based on existing spiking neural network models with STDP learning to show the computational feasibility of using such a mechanism to map the image of an object to its name with the same high accuracy as a state of the art machine learning classifier. 
Although not yet a theory, we believe this hypothesis presents a paradigm shift that is worth further investigating: it opens the way, among others, to new interpretations of language acquisition and understanding, the interplay between memories encoding and retrieval, as well as reconciling the apparently opposed views between sparse coding and distributed representations. 
\end{abstract}

\IEEEpeerreviewmaketitle
\section{Introduction}
Human brains process sensory visual input and learn to extract invariant representations from such input in an unsupervised manner. However, mapping \textit{signifiers}~\cite{saussure}, i.e. mental representations of the image-sounds or "words", to the \textit{signified} mental representations that such words refer to, needs an interaction with an external agent that supervises the learning. Such ``teacher" simultaneously generates the sound-image related to the signifier in the presence of the actual stimuli that relates to the signified: this can happen for example by generating the sound "cat" or the writing of the word "cat" in the presence of an actual image of a cat. The teacher shall repeat this procedure until both are associated. We say that the agent has learned to \textit{map the signifier to the signified}. 
 
In this context, it is tempting to think at first sight that the repeated co-occurrence of both stimuli reinforces their connection, ``in a hebbian manner'', thus allowing the mapping of signifiers to their signified representations. However, it is commonly accepted that neurons process sensory information mainly in a forward manner, i.e. from pre-synaptic to post-synaptic neurons.
Yet, in the case of the signifier and the signified ``problem'', the two co-occurring signals would eventually reach a connection point, where both signals have followed only forward paths. 
Now, assuming that memories are stored and encoded in the same areas\footnote{See the discussions in Sec.~\ref{sec:top:down} for more elaboration on this assumption} that uniquely responded to the memorized stimuli during the first encounter, a challenging question arises:  
What neural mechanism allows to associate one to the other, such that the activation of the signifier can trigger back the activation of the signified and vice versa.

We argue that this problem is a particular case of a more general one that occurs whenever the recollection of previously encountered and stored stimuli is needed. This is the case of explicit memory where sensory stimuli, such as a visual scene or a smell, act as a cue to trigger the retrieval of past related events. We further argue that this is also the case in imagination where different parts of previously encoded stimuli are recalled and ``merged" together to generate an ``imagined" new experience that was not exactly met before.

In this paper, we hypothesize that weak and fast-fading backpropagating action potentials (APs) (from post to pre-synaptic neurons) whose strength is proportional to pre-synaptic ``weights'' are the medium by which previously encoded information is recollected. Such signals start in what we call \textit{source pointer neurons} and travel all the way back, selectively reactivating on their path the neurons which uniquely responded to the stimuli during the first encounter, thus creating a similar experience to that of the first time. A source pointer neuron, as we will develop later is a neuron that specializes, thanks to past memorized co-occurrences, in selectively and invariantly~\footnote{As we will show later in Sec.~\ref{source:pointer:evidencce}, the existence of such selective neurons has been widely observed~\cite{quiroga2005invariant:anniston,connor2005friends}} responding only to the retrieval cue (e.g. signifier for language) or its associated memories to be retrieved (e.g. signified). We refer to this as \textit{the backpropagation-based recollection hypothesis} throughout the paper.

Prior work has extensively studied the interplay between visual perception and retrieval as happens in visual imagery (See for example Pearson's recent review~\cite{pearson2019human}). For example, in addition to the high overlap between areas involved in retrieval and perception, it has been observed-- thanks to Dynamical Causal Modeling (DCM) of activation patterns, that there exists indeed a reverse top-down signaling pathway, from higher-level to lower-level cortical areas, that is responsible for recollection of visual images~\cite{dijkstra2017distinct,dentico2014reversal}. However, following the traditional conception of forward propagation, these observed top-down activation patterns were attributed (wrongly, we argue) to backward recurrent feedback connections. 
Providing a biologically plausible computational model at the neuronal level that explains how backward recurrent connections (that use the pre to post synaptic path) can reactivate a previously encoded stimuli was beyond the scope of their work and remains to the best of our knowledge unsolved. For the sake of completeness, it is worth mentioning that recent years have actually seen the rise of "forward-based" computational generative models that come from machine learning and that can generate realistic images, the most notorious being Variational Autoencoders (VAEs)~\cite{VAE:1} and Generative Adversarial Nets (GANs)~\cite{GANs}. However, having been designed for a different purpose, it is not clear how they can  be put together to implement retrieval tasks, even in machine learning. Second and most importantly, their complexity and the supervised mechanisms they employ make them less likely to be biologically plausible~\cite{grossberg1987competitive:backprop,crick1989:backprop,whittington2019theories}.
We argue instead in this paper in favour of a simpler unsupervised mechanism where no local error information and no output target for supervision are needed: the same forward paths used for perception are simply used backward for retrieval.

After stating our hypothesis, in the general case (Sec.~\ref{sec:hyothesis:general}) and in the particular case of language understanding and naming (Sec.~\ref{sec:hypothesis:language}), 
we discuss its verifiability and run it against a thorough review of related literature (Sec.~\ref{related}). We found no evidence that allows to rule it out, but a significant body of experimental evidence that corroborates the plausibility of many of its assumptions. 
We find, for example, that fading away action potentials backpropagating to the dendrites have been widely measured and shown to exhibit interesting properties: they are stronger when the postsynaptic neuron is firing and most of all, they can be controlled by neuromodulation so as to increase their strength or disinhibit them (see Sec.~\ref{evidence}). We further review the literature about neural correlates of cue-based explicit memory retrieval, as well as the existence of sparse pointer neurons, and find further evidence supporting the framework of the hypothesis.

We then focus in the remainder of the paper on a particular case of our problem which is language understanding and particularly naming, which we computationally model (Sec.~\ref{sec:model}) and then simulate (Sec.~\ref{experimental}).
In this context, we define \textit{naming} as the act of retrieving the representation of the sound-image (signifier) that refers best to a presented visual stimuli (signified). We define \textit{understanding}, on the other hand, as the task of retrieving the signified representation that corresponds to a presented auditory or visual stimuli of a signifier. 
We leverage recent success~\cite{kheradpisheh2018stdp,mozafari2018first} in training Artificial Spiking Neural Networks (SNNs) with \textit{Spike Timing Dependent Plasticity} (STDP) learning to simulate a neural network that implements our hypothesis. We verify the computational efficiency of backpropagation-based recollection by comparing its accuracy in correctly naming a visual object, to that of a state of the art Machine learning algorithm. To further challenge the computational ability of our hypothesis, we test it on an extreme learning task, which is naming objects after seeing a single instance of each class. We find that backpropagation-based recollection leads on average and maximum to a higher accuracy compared to a Support Vector Machine (SVM) classifier. We are of course aware that the SNN models we use in this paper are not the brain, let alone from the particular implementation we leverage.~\footnote{We build on Perez's python implementation available on github~\cite{SDNN:implementation} from whom we obtain the authorization to use and modify for research purposes. We also release our modifications so as to ease the reproduction of results~\cite{Our:SDNN:implementation}} We believe, however, that the simulations hint towards the computational efficiency of the mechanism. That, especially tied with our literature review, calls for a serious further exploration of this path. Given the breadth of the potential implications (as we discuss in Sec.~\ref{sec:implications}), the backpropagation-based recollection hypothesis deserves at least to be rigorously and explicitly proven wrong through further studies.

\section{The backpropagation-based recollection Hypothesis}\label{sec:sec_rh}
\subsection{General case}\label{sec:hyothesis:general}
We posit our hypothesis and its assumptions as follows. 
\textit{When a stimuli is presented to a (``trained'') brain, the latter responds by a unique activation pattern, resulting from the activity of an ensemble of neurons that signal the various salient features of the stimuli. 
Memories of past encountered stimuli are stored in a distributed fashion by the same neural ensembles that detected such stimuli when encountered and encoded for the first time. 
Recollection of memories is therefore a process by which the appropriate population of neurons is re-activated again so as to "re-live", offline, an experience that is similar to the first encounter.
We hypothesize that weak and fading-away backpropagating signals from post-synaptic to pre-synaptic neurons, whose strength is proportional to the post-synaptic neuron's firing rate and pre-synaptic weights, is the mechanism by which the brain performs generative tasks. By generative tasks, we mean the regeneration of a previously lived and memorized stimuli (e.g. recollection in explicit declarative memory), the generation of a plausible future stimuli (e.g. future episodic memory) or the regeneration and combination of previous separately-lived and memorized stimuli, a process we refer to as imagination (e.g. imagining a cat that laughs by combining a memorized mental image of a cat with that of the act of laughing).} 

\textit{The recollection process starts by the presentation of a retrieval cue, which activates few sparse neurons that uniquely identify both the retrieval cue and the to-be retrieved memory. We assume that these neurons learned to respond only to the presence of either of both stimuli thanks to a low-level learning rule such as STDP~\cite{markram:stdp}; due to the (repeated or modulated) co-occurrence of both stimuli in the past: the cue and the to-be retrieved. We further hypothesize that the retrograde signal is initiated in such source "pointer" neurons (e.g. ``Jennifer Anniston'' cells~\cite{quiroga2005invariant:anniston} if the goal is to recall prior stimuli related to, say, Jennifer Anniston). The signal then travels backwards following the paths with higher weights, activating on its way all the various neurons that compose the mental images to be recalled.   
We hypothesize finally that such backpropagation can be controlled remotely via neuromodulation so as to invoke it, increase its strength or to inhibit it. This neuromodulation acts thus  as a ``switch" to control whether to retrieve or not.}

\subsection{Case of language acquisition, understanding and naming}
We further argue that a particular case of the above-mentioned tasks, which we later computationally simulate, is a form of explicit semantic memory related to language understanding and naming. We first start with some terminology.
\subsubsection{Terminology}
We adhere to the conceptualization and terminology introduced by swiss linguist Ferdinand de Saussure~\cite{saussure} and build on the distinction he introduced between the \textit{signifier} and the \textit{signified}. Since several interpretations of Saussure's work could perhaps be made, we clarify ours. We refer to the signifier as the mental representation of the sound-image of the word. By sound-image, we mean either the phonetic sound resulting from the word, or the image of the letters that form the word. We refer to the signified as the mental representation of the actual object that the sound-image and its mental concept refer to. Both the signifier and the signified are concepts, one represents the word, the other represents the mental image(s) that this word often refers to.
In this context, we refer to \textit{understanding} as the act of mapping the signifier to its signified. \textit{Naming} is the act of \textit{retrieving} the signifier that corresponds to a given mental representation or to a presented visual stimuli.

\subsubsection{Illustration in the case of language}\label{sec:hypothesis:language}

\begin{figure}[t]
    \centering
    \includegraphics[width=\linewidth]{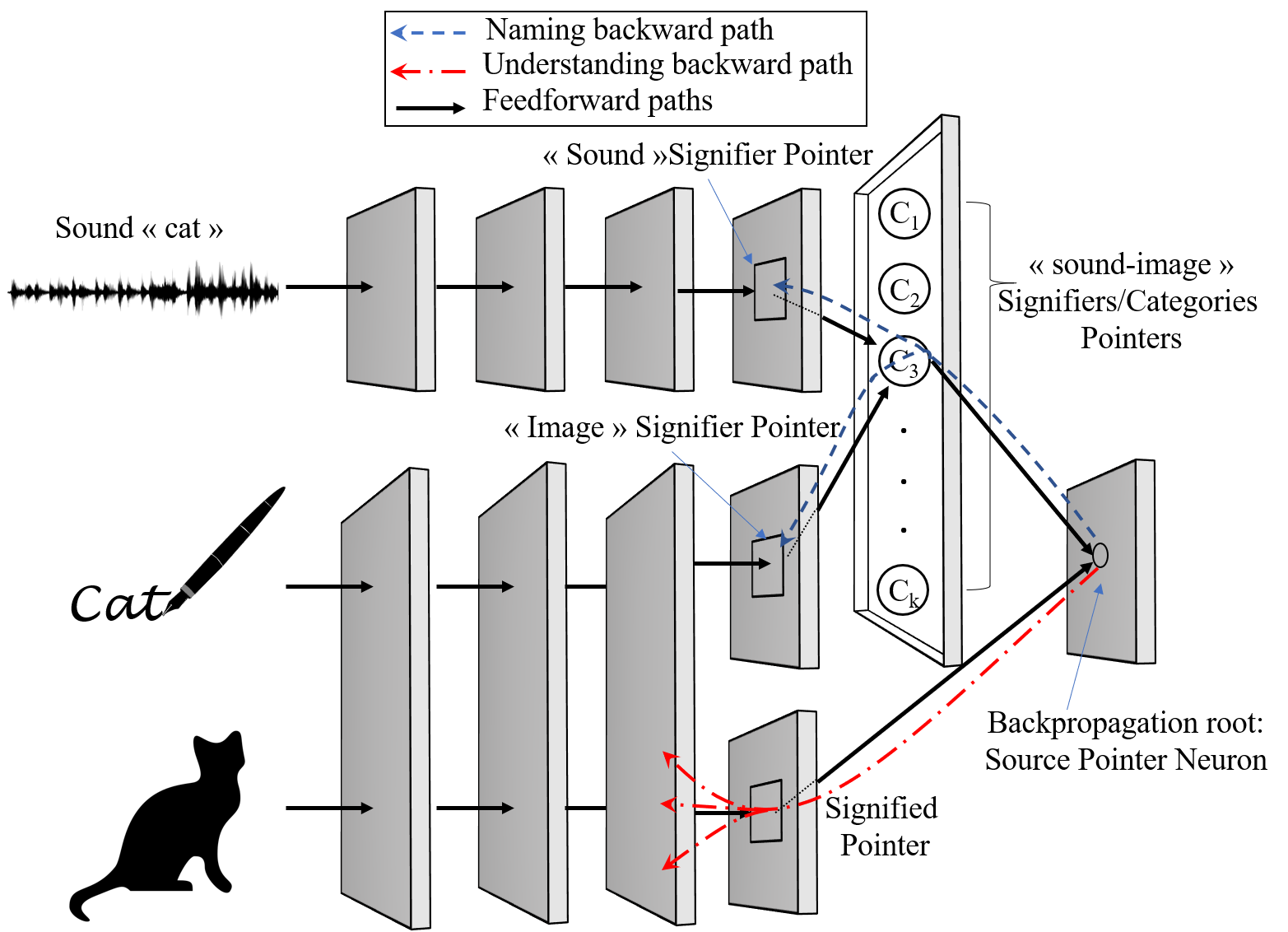}
    \caption{Illustration in the case of naming and understanding}
    \label{fig:hypothesis}
\end{figure}

When it comes to language, our hypothesis implies that backpropagating APs mediate understanding and naming. Fig.~\ref{fig:hypothesis} illustrates our modeling through the example of three concurrent sensory inputs that are presented to a learner and that need to be ``permanently associated''.
This exemplifies a ``teacher'' that shows the ``learner'' an image of a cat, simultaneously to how the word cat is written, as well as the sound of the word.
We assume as illustrated that there exists an area where the visual and sound activation pathways intersect (here, at the last ``backpropagation root'' layer). Such intersection areas are the root of backpropagating APs. During the ``understanding'' task, the retrieval cue is a word-related stimuli (e.g. sound ``cat'' or image of the word ``cat'') and the retrieved memory is the signified representation (illustrated by the Understanding backward pathway in the figure). In the task of naming, the retrieval cue is the signified object (here a cat) and the retrieved memory is the name or signifier of the object (illustrated by the Naming backward pathway).

In more details, the sensory input of \textit{a} sound ``cat'' as well as \textit{an} image of the word are processed through consecutive feed-forward neural layers. Similarly to what happens in primates' visual cortex~\cite{kruger2012deep:primate}, neurons in the earlier layers have learned (thanks to a simple unsupervised rule such as STDP) to respond to simple features and the deeper we go, the more selective the neurons become and the more they respond to more complex ones. We assume that at a later processing stage, there exists fewer ``sparse'' neurons that selectively respond only to the presence of the sound ``cat'', we refer to such neurons as the ``sound signifier'' pointer neurons of the word cat. Similarly, we assume that there exists neurons that selectively respond to the ``image'' of the word and we refer to such family of neurons as the ``Image signifier'' pointer neurons. Finally, neurons that selectively respond to both are referred to as the ``sound-image signifier'' or simply signifier pointer neurons. 
Following a similar pattern, the visual image of the cat itself is processed through various neural stages until certain neurons (referred to as signified in the figure) respond selectively only to the presence of the image of a cat.

This is how, as illustrated in the figure, the above described visual and sound pathways reach a common connection point, in a subsequent feedforward neural layer. We hypothesize that the repeated, or the neuromodulated\footnote{The connection can be reinforced either by mere repetition, e.g. repeating many times the word cat in the presence of an image cat, or by neuromodulation, in which case a single co-occurrence can be enough.}, co-occurrence of signifier and signified stimuli (i.e. saying ``cat'' in the presence of an actual cat) reinforces their connection, at this junction point layer, in ``a hebbian manner'' thanks to a simple rule such as STDP learning. Next, when presented with either signifier or signified related stimuli, the \textit{source pointer neuron(s)\footnote{one in theory should be enough but there must be many in reality, e.g. at least for redundancy to prevent failures} in the backpropagation root layer will fire, resulting in backpropagating action potential(s) that are proportional to pre-synaptic weights}. The latter will cause the backward activation of the appropriate neurons, thus recalling the signified if the presented stimuli relates to the signifier, and vice versa.

It goes thus without saying that we adhere to the view that there exists few neurons that respond selectively to complex stimuli such as (i) sound signifier stimuli, (ii) image signifier stimuli, (iii) signified or (iv) uniquely to any of the three previous ones. During backpropagation-based retrieval, these neurons act as \textit{pointers} to selectively reactivate an appropriate population of pre-synaptic neurons that uniquely characterizes the memory trace to be retrieved; thus creating a similar experience to that of the first encounter(s)\footnote{For simplicity, we focus in this toy example, on a single encounter that, we assume, was ``encoded right away''. In reality, stored memory traces might evolve with repeated exposure, such that recalling what is meant by the word cat, leads to the recall of a memory of the signified, that is statistical in nature (e.g. one of the many cats met before, or an average abstract image of a cat)} when the memory was encoded. For example, what is retrieved could be an experience of the sound of the word "cat" with the particular voice or conditions in which it was encoded (in line with what is called the encoding specificity principle~\cite{tulving1973encoding}).  

This is how, as we will discuss later, \textit{our hypothesis reconciles (i) sparse/localist and (ii) distributed representation theories in the brain, promising to end a long debate between cognitive psychologists and neuroscientists~\cite{bowers2009biological,grandmother:psycho:neuro}}. 
In our framework, there is no need to chose between them as both are needed, but for different purposes: Sparse coding, illustrated here by the presence of highly selective neurons, is needed for backpropagation-based retrieval, while the encoding of the entire memory trace is still done via a distributed set of neurons. The latter can be selectively reactivated ``on-demand'', from source pointer neurons backwards. It is thus the simultaneous activation of an entire population of neurons that forms the entire memory trace, and single highly selective neurons are only pointers, helpful for retrieval. 

Finally, and anecdotally, \textit{the fact that backpropagating signals are ``fading-away'' in nature could explain the ephemeral nature of the experience of recalled memories or visual mental imagery's lack of vividness}: the latter are not as persistent as the experience of live sensory stimulation.


\subsection{What this hypothesis is not about}\label{sec:hyothesis:notabout}
Finally, we clarify that this hypothesis covers \textit{only the recollection processes}. This means, in the particular case of explicit recall, the 
\textit{reactivation, as close as possible, of the same neurons} that were activated during previous encounters.
As a consequence, in the case of language, what is covered by our hypothesis is how to (i) learn the name association and how to (ii) recall the name, \textit{not} yet actually how to \textit{produce} it (e.g. emitting the sound or writing the letters). Further investigations are needed to reassess production tasks in light of our new hypothesis. Nonetheless, it occurs to us that learning to produce the right sounds, i.e. speaking, happens through a trial and error, reward-based based process in which the goal is to ``mimic''. The study of such mechanism is beyond the scope of this paper. 
Our hypothesis covers how the reactivation of source pointer neurons leads to recollection. How such pointers further participate to invoke motor areas to produce sounds or write letters is left for future work.

Next, when we talk about understanding, we mean the modality-specific features of semantic memory~\cite{patterson2007you}:  i.e. recalling the details of how a face or an emotion look or feel like exactly, as opposed to other aspects of semantic memory like finding abstract relationships between words. We leave the latter aspect also for future work.  Worth mentioning though, Patterson \textit{et al.} reviewed semantic knowledge organization in the human brain~\cite{patterson2007you} and reported that all theories agreed on the fact that modality-specific recall is implemented by a distributed brain network, a fact that is coherent with a backpropagation-based recollection hypothesis. 

Then, our hypothesis assumes that there are centers that remotely control via neuromodulation whether or not to call the recollection (by either increasing the backpropagation or inhibiting it). Our hypothesis does not cover what mechanisms control these control centers, and under which conditions recollection is favoured or shut down. What our hypothesis predicts is that the task of these ``control centers'' can be extremely easy to implement: the diffuse and \textit{``untargeted''} remote generation of an excitatory neuromodulator favours further recall (i.e. recall of whatever cues are active or firing at the moment). The same applies for inhibition.

Finally, our hypothesis, being only focused on the recollection process from sparse pointer neurons backwards, does not directly explain the mechanisms involved in sparse neurons formation, novelty or familiarity detection, and the interplay between short (e.g. few days back) and long (e.g. few years) term memories. Nonetheless, it still offers a ground to reason about these issues. For example, the fact that the same sparse neurons keep being used to signal the presence of the same familiar stimuli throughout the years (e.g. face of own child) could explain why humans are unable to remember much younger versions of these faces (in the absence of photos): memories are updated \textit{in situ} and familiar faces will always lead to the activation of the very same ``familiar'' invariant sparse neurons, not to the activation of ``novel'' ones.

\section{Review of evidence in the literature}\label{related}
In this section, we position our hypothesis in the literature and show evidence that backs it up.

\subsection{Existence of backpropagating action potentials}\label{evidence}
Despite the forward-processing prevalent view, a plethora of studies have measured, \textit{in vitro} and \textit{in vivo} in anesthetized~\cite{svoboda1997vivo:anesthesized} and awake~\cite{bereshpolova2007:awake,1998somadendritic:awake} mammalians, action potentials that backpropagate to apical and distal dendrites, and this for various classes of neurons~\cite{stuart1997action,vetter2001propagation,waters2005backpropagating,williams2000action:TC,williams2000backpropagation}. 
We cite in the following some of them. For a more complete list, the reader can refer to the reviews of Stuart \textit{et al.}~\cite{stuart1997action} or Waters \textit{et al.}~\cite{waters2005backpropagating} which summarized the findings about the measurements and hypothesized few roles of backpropagated action potentials. For a more general review of retrograde signals, i.e. not only activity-dependent but also during synaptogenesis etc., the reader can refer to Tao \textit{et al.}~\cite{tao2001retrograde}.

Williams and Stuart~\cite{williams2000action:TC} performed simultaneous somatic and dendritic recordings from Thalamocortical (TC) neurons and measured that action potentials, both due to sensory information or cortical excitatory postsynaptic neurons, backpropagate into the dendrites. In another work, the same authors\cite{williams2000backpropagation} measured the same phenomenon in neocortical pyramidal neurons. Interestingly for our hypothesis, the authors have found that action potentials due to physiological patterns of firing, backpropagate three to four times more effectively compared to action potential pertaining to mean firing rates. This observation is confirmed by several studies (reviewed by Waters \textit{et al.}~\cite{waters2005backpropagating}) which found that backpropagation was modulated by synaptic input. For example, properly timed excitatory input leads to the amplification of backpropagation, whereas inhibitory input might block it. More interestingly for our hypothesis, many neuromodulators have an influence on backpropagation, often leading to its enhancement, but in more complex ways. 
For example and interestingly, given the supposed role of the hippocampus in retrieval (see later), in hippocampal CA1 pyramidal neurons, it has been observed that muscarinic agonists enhance the backpropagation in a progressive manner having a stronger and stronger effect on subsequent action potentials~\cite{tsubokawa1997muscarinic:neuromod,hoffman1999neuromodulation}.
This suggests that neuromodulation therein can act as a ``switch" to enable Action Potential backpropagation in a selective manner, a feature that is necessary for our hypothesis. As described in Sec.~\ref{sec:sec_rh}, not every presentation of a visual stimuli would systematically lead to the activation of ``naming". And probably not every presentation of the signifier stimuli should lead to the evocation of its signified representation. Similarly, not every exposure to a familiar stimuli (\textit{known}) automatically leads to explicitly \textit{remembering} its context. 

Nonetheless, to the best of our knowledge, the role of such activity-dependent backpropagation of Action Potentials, as reviewed for example by Waters \textit{et al.}~\cite{waters2005backpropagating}, has been so far hypothesized to be local, acting as a feedback loop from postsynaptic to pre-synaptic neurons, to regulate spiking activity or support synaptic plasticity. In this work, we emit the hypothesis that such retrograde signals play a more explicit and direct role in higher-level cognitive tasks, such as naming, understanding, and other generative processes like imagination and explicit memory retrieval. We next oppose our hypothesis to the state of knowledge in experimental cognitive sciences about explicit memory (Sec.~\ref{sec:memory}). We later dive deeper in details and oppose our assumptions to what is known about the neural correlates of explicit memory(Sec.~\ref{sec:correlates:memory}).


\subsection{Cue-based retrieval: a cognitive sciences perspective}\label{sec:memory}
Our hypothesis applies to any task in which the reconstruction of previous encoded stimuli is needed. The literature on this topic found its origins in early experimental cognitive science research (e.g. \cite{tulving1966availability,tulving1972episodic}) before recent advances, driven by neuroimaging and optogenetic stimulation, allowed to shed more and more light on some actual neural correlates~\cite{frankland2019neurobiological}. We start by surveying the first and linking it to our hypothesis.

Since the seminal work of Endel Tulving, long-term human memory is widely classified into explicit (declarative) memory and implicit (procedural) memory. While procedural memory relates to long-term acquired skills such as driving or playing an instrument, declarative memory, relates to the explicit recollection of memories about facts, words, images and events etc. In this paper, we focus on the latter and provide a hypothesis about how the retrieval of these memories happens at the neuronal level. It is worth mentioning that Tulving also played a role in further dividing explicit memories into episodic and semantic ones\cite{tulving1972episodic}. Our hypothesis is orthogonal to the difference between them.

One crucial principle in this area was formulated by Tulving and Thomson under the name of encoding specificity~\cite{tulving1973encoding}. The principle stresses the importance of retrieval cues and the entire context that is perceived during encoding for later retrieval:  the surrounding context that was present during the first perception and encoding moment can act as an efficient retrieval cue in the future. 
Although it might sound straightforward today, such early work~\cite{tulving1966availability,tulving1973encoding} played a role in differentiating between \textit{availability} of memories and their \textit{accessibility}, \textit{thanks to cues, let them be internal or external stimuli-based}: the inability to recall does not necessarily mean that the memory is not available but could be also due to the lack (inactivation) of appropriate cues. 

Our hypothesis offers a neurobiological ground to interpret and simulate encoding specificity: for us, any surrounding context during encoding can act as a retrieval cue as long as it activates the source pointer neurons that uniquely\footnote{The selectivity of source pointer neurons is a consequence of encountered prior stimuli, and this defines what particular contextual cues are efficient at recalling what particular encoded trace.} identify the retrieval cue and all the remaining surrounding context to be retrieved. Backpropagating action potentials can then ``travel backwards" to reactivate networks of neurons which uniquely responded to the stimuli during the encoding moment, thus creating, again, a similar experience.

In fact, the process by which an internal or external (sensory) cue activates the stored memory trace is well known and has been called ecphory~\cite{tulving1983ecphoric,schacter1978richard}. Ecphory\footnote{A term that Tulving revived together with the forgotten work of german scientist Richard Semon, who first coined it and stressed the importance of retrieval cues.}, which is this interaction between trace and cue, is described as the first stage of memory retrieval before conversion actually happens and the recollection experience is ``lived''. In our hypothesis, the cue activates mainly the source pointer neurons, which in turn activate back the appropriate presynaptic populations, resulting in the recall of all the related traces.
\textit{As such, following our hypothesis, it can be easily seen how two components may affect memory retrieval as already predicted by Tulving: the lack of the appropriate cue or a decay of the synaptic weights of the concerned neural networks. Our hypothesis announces others that could be verified. For example, a decline in the intensity or extent of the backpropagation (e.g. impairment in the neuromodulation that is supposed to facilitate it) can hamper retrieval. Conversely, an excess of such backpropagation might lead to higher levels of intrusive thoughts.}

Today, the encoding specificity view has passed the test of time (see  ~\cite{frankland2019neurobiological} for a recent review) and even its ``opponents''~\cite{nairne2002myth,poirier2012memory,goh2012testing} do not question the necessity of some degree of match between encoding and retrieval conditions, but rather stress the importance of additional factors that influence the performance of later retrieval; the most important being the ``discriminative'' power of the retrieval cue or its distinctiveness. Accordingly, recalling performance is not only related to the amount of match between encoding and retrieval conditions, as thought first, but also to cue overload and hence to what extent the cue is ``discriminative''.
\textit{This latter view can be easily observed under our framework: let's have in mind the illustration in Fig.\ref{fig:hypothesis} and the image of a cat as a cue. If this image appears, during learning, simultaneously with all sorts of names, and  not only the signifier ``cat'', then the backpropagating action potentials would potentially simultaneously activate many `` Signifier" neurons (and not only that of the cat), making it hard to distinguish and hence correctly name. The necessity of being discriminative can be seen in our simulations of naming later in Sec.~\ref{experimental}: after backpropagating action potentials back to the ``categories'' layer, the ``signifier neuron'' that gets the highest ``votes'' is elected to signal the name of the object. If all neurons receive ``equal votes'' because of cue overload, it would be impossible to retrieve the right name.}  

\subsection{Explicit memory: neural correlates}\label{sec:correlates:memory}
We now dive more into the neurobiological foundations of encoding specificity, ecphory and the processes involved in explicit memory. This field has seen tremendous advances driven by both neuroimaging and optogenetic stimulation. 

We start from a recent thorough review~\cite{frankland2019neurobiological} in which a large body of research strongly supported Semon's and Tulving's cognitive theories: namely, that (i) the success of accessibility depends on the interaction between cues and memory traces and that (ii) there are strong ties between encoding and retrieval, including at the level of activated neural ensembles. In particular, it was shown, using artificial optogenetic stimulation\footnote{A technique that allows to later activate or inhibit precisely only selected populations of neurons that were initially selectively tagged, depending on their activity} that it is possible to either disrupt or mimic ecphoric processes by activating or inhibiting the \textit{same specific neural ensembles} that were active during encoding\footnote{Note that, here, we use interchangeably encoding, learning and conditioning to accommodate different terminologies used in different papers.}. 

In one experimental study~\cite{tanaka2014cortical} that we further analyze below, blocking the neural ensembles that were used during encoding to recognize the cue, resulted in impairment in retrieval. In the experiments, mice were conditioned to produce a fear response whenever placed in a particular context, a context which stands here for the cue. At the same time, CA1 neural ensembles that were particularly active during encoding are optogenetically tagged. Whenever placed in the same context again, mice successfully freeze as a sign that they recognize the environment. However, placing them while inhibiting the previously tagged neural ensembles considerably reduces freezing levels. \textit{This means that if the neural ensembles that recognize the cue do not activate, the memory is not retrieved.} Other studies (e.g.~\cite{liu2012optogenetic}) similarly used optogenetics to demonstrate the ``opposite'': artificially reactivating the neural ensembles that recognize the cue, thus inducing the retrieval of the memory trace (i.e. causing freezing), even outside the context in which the conditioning happened (i.e. in the absence of a natural cue). Now, in the previous two families of experiments, (i) mapping the cue and the trace was learned naturally, (ii) neural ensembles were tagged depending on their activity during encoding, and (iii) artificial inhibition or excitation was used to disrupt or elicit retrieval. A last recent family of experiments~\cite{vetere2019memory} showed that it is even possible to associate a cue and a trace artificially and to later elicit retrieval in natural conditions. In particular, they repeatedly used photo-stimulation to artificially activate a neural ensemble that usually recognizes a special smell, simultaneously to photo-stimulating another memory trace that elicits avoidance. After this co-occurrence based conditioning happened, exposure of mice to the real smell caused an avoidance reaction: mice ``remembered'' to avoid although they've never experienced the smell in real conditions. 

Finally, note how, in optogenetic stimulation above, light simultaneously and selectively activates a set of neurons that were prepared in advance to become light sensitive. This allowed scientists to dissect the interactions between memory traces, retrieval cues and encoding. However, it is not clear today what mechanism in the brain can cause a similar selective reactivation, equivalent to optogenetic stimulation. \textit{That's exactly the purpose of our hypothesis: hence, our answer to the question ``how does a neural ensemble activates in a selective way'' lies in backpropagating action potentials following the paths with highest presynaptic weights.  Our hypothesis also offers a framework to  simulate (as we do later) and understand these issues at the level of single neurons.}

As a first summary, the studies above~\cite{tanaka2014cortical,liu2012optogenetic,vetere2019memory} and many others reported in the review~\cite{frankland2019neurobiological} (which we encourage the reader to check) confirm the importance of cues and encoding specificity. But beyond that, they suggest that retrieval reactivates what was active during encoding, in a process referred to, sometimes, as neural reinstatement. \textit{This principle is at the heart of our hypothesis as backpropagated APs should follow exactly the reverse path that uniquely led to the activation of source neurons during encoding. It turns out that many arguments support this reinstatement principle. We review them in what follows.}

\subsubsection{Retrieval as top-down re-activation}\label{sec:top:down}
First, historically, the oldest (yet weak) supporting fact is that simply reinstating the encoding context during the recollection time enhances retrieval performance and quality, as reported by some reviews~\cite{smith2001environmental,eich1995mood}. 
Second, more recently and more strongly, a significant body of research intentionally studied the overlap between encoding and retrieval and provided large evidence in favour of the principle using various ensemble tagging\cite{liu2012optogenetic,tanaka2014cortical,denny2014hippocampal,reijmers2007localization,sorensen2016robust,lacagnina2019distinct,khalaf2018reactivation,ramirez2013creating,tayler2013reactivation,guskjolen2018recovery}, EEG\cite{waldhauser2016episodic} and neuroimaging\cite{jafarpour2014replay,johnson2009recollection,manning2011oscillatory,ritchey2013neural,staresina2012episodic,yaffe2014reinstatement,fulford2018neural, dijkstra2017vividness, dijkstra2019shared,dijkstra2018differential,dijkstra2017distinct} techniques. 
Furthermore, it was shown that the reactivation overlap between encoding and retrieval influences also the perceived quality of the retrieval. For example, in the case of visual imagery (i.e. attempting to mentally visualize an image), it has been shown that activation overlap in visual cortex increased visual imagery vividness, or the subjective intensity of the remembered image\cite{st2015distributed, dijkstra2017vividness,fulford2018neural}.
We refer the reader to the references above for more information and cite in what follows only few examples of each technique. 

In terms of neuroimaging, Dijkstra \textit{et al.}~\cite{dijkstra2017distinct} used Dynamical Causal Modeling (DCM)~\cite{friston2003dynamic} to infer coupling between cortical regions involved in the tasks of visual perception as opposed to visual imagery. They measured that visual imagery vividness correlated more with top-down connectivity patterns (from high level cortical areas to lower level areas) as opposed to perception itself. Many other studies suggest such top down mechanism during visual imagery~\cite{hochstein2002view,serre2007feedforward,linde2019evidence,dentico2014reversal,pearson2019human}. The reader can refer to Pearson's recent review~\cite{pearson2019human} of the cognitive neuroscience of visual mental imagery for more details about the top down reverse hierarchy of information and the fact that the process seems to be a weak form of the bottom-up perception. However, in general, due to the widespread view that neural computation is mainly forward from pre to postsynaptic neurons, this top down activation cascade has been always interpreted, in the literature, as the result of feedforward feedback connections from higher-level cortical layers to lower ones. 

Then, and perhaps more convincing than neuroimaging, neural ensemble tagging techniques also confirm the principle. In addition to work we described above~\cite{tanaka2014cortical,liu2012optogenetic,vetere2019memory}, another recent example is the work of Guskjolen \textit{et al.}~\cite{guskjolen2018recovery} who performed contextual fear conditioning experiments on young mice while tagging the neural ensembles which were active during encoding. As happens with infantile amnesia in humans, the infant mice later exhibited forgetfulness. However, photo stimulation of tagged neurons, only in the hippocampal formation (Dentate Gyrus in particular), induced memory recovery and reactivation of broader areas which were tagged during conditioning including hippocampal CA1 and C3, and cortical neurons. This finding is inline again with the idea that traces are distributed in neural ensembles that span many cortical brain regions \cite{wheeler2013identification}, each responsible of one aspect of information (sensory, motor, visual, emotional etc).

\subsubsection{Existence of ``backpropagation root layers''}\label{root:layer}
The last paper leads us to the last point we review in this section: the role of the Medial Temporal Lobe and its relationship to our backpropagation root layers, where source pointer neurons are located. \textit{Indeed, our hypothesis assumes the existence of an area where source pointer neurons lie and where the backpropagation starts. If our hypothesis is correct, this area should form the ``glue'' between cues and retrieved traces, and should observe a reversal of the flow of information. Interestingly, the Medial temporal lobe and the hippocampus in particular has been shown to (i) play this role and (ii) exhibit a similar reversal behaviour.}

For (i), many theories\cite{squire1995retrograde,teyler2007hippocampal,teyler1986hippocampal,mcclelland1995there,merkow2015human} support that the hippocampus performs exactly the task of reinstating patterns of activity in the cortex that were alive during encoding. 
This can be already seen from the study of Tanaka \textit{et al.}\cite{tanaka2014cortical} which we reported above. What they did by actually monitoring cortical activity while inactivating hippocampal CA1 cells on rodents, shows how the hippocampus is likely responsible for reinstating the patterns that were active at encoding. By permanently tagging neurons which were active during encoding (in a fear conditioning experiment), they were able to silence them with laser stimulation, up to several days later. When silencing only the tagged CA1 cells (and not the entire engram or trace ensemble), memory retrieval was impaired; and the rest of the neural ensemble in the cortex and amygdala, which used to reactivate during retrieval, was not reactivated again. 
Many other studies~\cite{danker2017trial,horner2015evidence,staresina2013reversible} also showed that retrieval success depended on whether or not the hippocampus was concurrently solicited or not, and this during both encoding and retrieval. 
Horner \textit{et al.}\cite{horner2015evidence} showed further evidence that the hippocampus binds together all elements composing a trace that is stored in distributed regions in the cortex, playing as a hub to perform what is also called pattern completion task. Finally, interestingly for our hypothesis, Staresina \textit{et al.}\cite{staresina2013reversible} observed a reversible signal flow from the cue region to the ``to be recalled'' target region, through the hippocampus. \textit{This puts the HC and MTL in the position of good candidates to be root backpropagation areas as per in our hypothesis: they seem to implement the link between cues' networks and traces' networks and they seem to be the place where flow reversal happens.}

Citing verbatim a thorough review and perspective from Moscovitch\cite{moscovitch2008hippocampus}: ``Retrieval occurs when an external or internally generated cue triggers the hippocampal index, which in turn activates the entire neocortical ensemble associated with it. In this way, we recover not only the content of an event but the consciousness that accompanied our experience of it''. Moscovitch refers later to the hippocampal memory indexing theory of Teyler and DiScenna~\cite{teyler1986hippocampal,teyler2007hippocampal} as follows: ``Memory traces in the HC/MTL are encoded in sparse, distributed representations that act as an index or pointers to the neocortical ensembles that mediate the attended information''. \textit{This claim, which is inline with our hypothesis, leads to the next assumption which we further oppose to the literature: the existence of sparse source pointer neurons.}

\subsection{Existence of highly selective source Pointer Neurons}\label{source:pointer:evidencce}
We assume that, at a certain deep level of processing, certain neurons will become highly selective and invariant: they serve, in our hypothesis, as pointers to reconstruct the encoded stimuli they represent. For example, in the particular case of language, some neurons will respond only to the signified or only to the signifier (one of the cues), and some would respond to both the signifier and the signified (i.e. what is in common between the cue and the ``to be recalled''). In this section, we scan the literature about concept representation in the brain to assess the plausibility of the existence of such neurons. \textit{It turns out that similar neurons have been documented and that their role is not yet well understood, given the still ongoing debates between two opposed views on the matter: the ``distributed representations'' view~\cite{georgopoulos1986neuronal,decharms2000neural} and the `` sparse coding'' view~\cite{barlow1972single,olshausen2004sparse}}. 

Indeed, the distributed representations view defends that concepts in the brain are represented by the unique activation patterns of entire and large populations of neurons. It is thus the pattern uniqueness across a large population that defines complex concepts, not particular single neurons. The sparse coding view defends instead that there exists few neurons that represent selectively particular items or concepts. The extreme version of sparseness would be that there is a unique cell that responds to each single unique concept, a version that is pejoratively known as the grandmother cell hypothesis~\cite{bowers2009biological,grandmother:psycho:neuro}. \textit{Our hypothesis promises to reconcile both views as follows: information is stored in distributed networks, and sparse neurons, which also exist, play the role of hubs to connect such distributed networks, easing retrieval by being sources of backpropagated APs.} 

Now looking at the literature, the first accounts of sparseness are quite old already. In practice, since the seminal work of Hubel and wiesel~\cite{hubel1962receptive}, it became mainstream that neurons tend overall to respond to more and more complex features, the deeper we go in the processing layers of sensory input\cite{kruger2012deep:primate,mishkin1983object}. Indeed, evidence suggests the existence of a hierarchy starting from the primary visual cortex V1, where basic features are encoded, until the inferior temporal (IT) cortex, where neurons selectively respond to complex shapes like hands and faces~\cite{gross1969visual,tanaka1996inferotemporal,logothetis1996visual}. 

Other known examples of sparse coding, for spatial representation, are place and grid cells~\cite{moser2008place}. Place cells for instance are single neurons which signal specific places in the environment: as the individual navigates in its environment, only the neurons that signal the current place field fire. Interestingly, such highly selective neurons have been found within the hippocampal formation, the area which seems to be a good candidate for being a backpropagation root layer as discussed in Sec.~\ref{root:layer} above.

In general, the literature is rife with studies that have measured such selective neurons, in ways that fit our hypothesis, and interestingly in these same MTL areas. For example, Fried \textit{et al.}~\cite{fried1997single} have measured neurons that selectively discriminated humans (faces) from inanimate objects, and this, interestingly, during both encoding and retrieval. Others distinguished specific facial expressions. A little later, Kreiman \textit{et al.}\cite{kreiman2000category}  have measured neurons that highly responded only to specific categories such as animals, houses and celebrities.   

In a continuous line of work, Quiroga and  colleagues~\cite{quiroga2005invariant:anniston,quiroga2008sparsebutnot,waydo2006sparse,quiroga2014single} have set to understand how the visual features we mentioned above are passed to upper layers of the hierarchy so as to understand how they are later used by higher cognitive processes: a question to which we hypothesize an answer in this paper. 
It is in one of these works, which became popular, that Quiroga \textit{et al.}~\cite{quiroga2005invariant:anniston} reported the existence of highly selective neurons that responded to the presence of specific stimuli related to places or individuals such as Bill Clinton and Jennifer Anniston. One of the found selective neurons even exhibited highly selective responses to any stimuli that is related to Halle Berry, let it be the face, or even the written words. \textit{The latter neuron exhibits striking similar properties to our Source pointer neurons as described in the language understanding and naming tasks.} Then, given that this work reminds the widely unaccepted grandmother cell hypothesis, further clarification have followed up.

Waydo \textit{et al.}~\cite{waydo2006sparse}, with Quiroga as a co-author, later used a probabilistic approach to explore a bit more rigorously Quiroga \textit{et al.}'s original findings~\cite{quiroga2005invariant:anniston}. Indeed, the latter obviously did not test for all MTL neurons and all possible categories of objects. Hence, (i) a found selective invariant neuron could respond to other untested categories, and (ii) there might exist many neurons, and not only one as found by the authors, that would selectively respond to the same stimulus. Authors thus developed a probabilistic model to estimate the odds, and confirmed the sparseness hypothesis (yet, arguing also against single grandmother cells~\cite{barlow1972single} as done first by Quiroga and co-authors). Nonetheless, the model leads to, as the authors conclude, only a bound on the true sparseness: the neural coding could be in reality even much sparser than they estimated. 
In another follow up work, Quiroga \textit{et al.}\cite{quiroga2008sparsebutnot} insisted, already in the title, on the fact that it is sparse but not grandmother cells and argued against the unlikely possibility that a single unique neuron responds to each stimulus. \textit{In our case, although we simulate the hypothesis using single neurons in Sec.~\ref{sec:model}, our hypothesis is inline with existence of many of such sparse invariant neurons. On the contrary, we believe that multiple neurons shall be needed, at least to guarantee resiliency if some neurons fail.} The authors however conclude with a set of difficult open questions. Our hypothesis suggests already few answers to the following ones: ``How are MTL cells involved in learning associations? How are MTL cells involved in free recall or the spontaneous emergence of recollection in the human mind?'' As discussed, we hypothesize it is: backpropagated APs, triggered through neuromodulation by remote ``control centers'' that simply decide whether or not to facilitate the recall, the naming etc. The same principle should apply to free recall, where said centers activate sequentially related concepts (see our discussion on mind wandering in Sec.~\ref{sec:implications}). 
 
Last but not least, in a subsequent work~\cite{quiroga2014single}, Quiroga \textit{et al.} have measured that the MTL selective neurons reflect the subjects' decisions about the stimuli rather than visual features themselves. They've put this in evidence by performing experiments in which they present subjects with vague stimuli that is a mixture of different celebrities (e.g. a picture that is the mixture of presidents Bush and Clinton). As expected from previous studies, exposing users to one of the celebrities first, leads them to later see the morphed image as pertaining to the other celebrity. This is probably due to the fact of ``tiring'' the neurons that encode this celebrity. Later by recording Clinton's and Bush's neurons, they concluded that such MTL neurons fire in accordance to the decision, not the features. 
In an interesting follow up comment, Reddy \textit{et al.}\cite{reddy2014concept} remarked that damages to the MTL area cause subjects to have memory impairments, yet having perfect perceptual awareness and consciousness. \textit{This is inline with our hypothesized role of source pointer neurons that allow to map related stimuli with backpropagated APs.}

As a summary, sparse neurons that respond selectively to complex concepts exist, in areas where we suspect them to do, with properties that are inline with our backpropagation-based recollection hypothesis. 

\subsection{Summary of arguments in favour}\label{summary:arguments}
We now summarize, as simplified in Table.~\ref{tab:supporting}, the arguments in favour of our hypothesis. 
First, as seen in Sec.~\ref{evidence}, activity-dependent backpropagating action potentials are biologically plausible. Moreover, it has been observed that these APs are stronger when neurons are firing, a necessary condition for our hypothesis. Second, neuromodulation can enhance such backpropagation in a selective and progressive manner, thus acting as a switch to inhibit it or to strengthen it. 
This ``feature'' is necessary as humans can control whether or not to favour the retrieval of some explicit memory after the exposure to a cue. Indeed, not each time we see a screen of laptop, we recall its name, or other related memories.  
Interestingly enough, this modulatory phenomenon on backpropagated APs has been observed in hippocampal cells, an area that is known to be crucial~\cite{merkow2015human,moscovitch2008hippocampus} in memory retrieval, and especially known in some theories~\cite{teyler1986hippocampal,teyler2007hippocampal,moscovitch2008hippocampus} as the place that stores the indexes that allow to retrieve memories that are stored in other cortical areas. We found also evidence that such sparse indexes which we called pointer neurons have been also observed and well documented~\cite{quiroga2005invariant:anniston,quiroga2008sparsebutnot}. Even more interestingly, we've seen that a reversal of information flows has been observed in the hippocampus which is believed to act as a glue between the cues and the engrams to be retrieved. 
This brings us to another assumption of our hypothesis which is that retrieval is the reactivation of the same areas and neurons that were used during encoding, a task sometimes called neural reinstatement, or pattern completion. We find a large body of optogenetic-based and neuroimaging-based evidence that confirms such assumption. There is indeed a high overlap between the areas involved in these two tasks, and optogenetic-based experiments cited above are indeed based on tagging the specific neural ensembles that were active during encoding. 
Additionally, and perhaps as an added bonus, we will computationally show later that this hypothesis is an effective computational method to associate names to visual input, with the same high accuracy of a supervised machine learning algorithm.

Additionally and may be anecdotally, the fact that these signals are weak and fading away might explain why imagination and memory recollection elicit subjective experiences that are themselves transient and fading away in nature: the recollection of an image of a cat is much less vivid and persistent than the subjective experience that is due to the sensory input of a real cat.

Finally, not reviewed in details above, if our hypothesis proves true for cue-based recollection, it becomes more than reasonable to embrace the view that it also mediates other generative tasks such as mind wandering, intentional creative thinking, dreaming, as well as future episodic thinking or imagining the future. Existing neural correlates studies of such generative tasks~\cite{christoff2016mind,kucyi2018just,addis2007remembering} can be leveraged to further verify our hypothesis.

To summarize, we interpret all the arguments in favour as an encouraging call for future work and further investigations. In particular, it should be verifiable experimentally if the extent of the action potential backpropagation is proportional to pre-synaptic weights, and whether and to what extent backpropagation can be far reaching (e.g. eventually one or more pre-synaptic hop(s) away). The latter is a crucial point that can drastically increase or decrease the plausibility of our hypothesis, and for which we did not find evidence neither in favour nor against. 
\begin{table}[]
\begin{tabular}{l|l}
Assumption                                          &  Evidence                \\ \hline
Backpropagating Action Potentials                   & \multicolumn{1}{c}{\checkmark} Sec.\ref{evidence} \\ \hline
Backpropagation stronger when neurons fire          & \multicolumn{1}{c}{\checkmark} Sec.\ref{evidence} \\ \hline
Backpropgation can be selectively modulated         & \multicolumn{1}{c}{\checkmark} Sec.\ref{evidence} \\ \hline
High overlap between retrieval and encoding         & \multicolumn{1}{c}{\checkmark} Sec.\ref{sec:top:down} \\ \hline
Information flow reversal (at pointer neurons)         & \multicolumn{1}{c}{\checkmark} Sec.\ref{root:layer} \\ \hline
Backpropagation effects can be far reaching         & No, but verifiable         \\ \hline
Backpropagation proportional to presynaptic weights & No, but verifiable         \\ \hline
Existence of source pointer neurons                 & \multicolumn{1}{c}{\checkmark} Sec.\ref{source:pointer:evidencce}
\end{tabular}
\caption{Summary of assumptions and supporting evidence}\label{tab:supporting}
\end{table}

\section{Name association: modeling with spiking neural networks}\label{sec:model}
We now focus on the task of retrieving object names using their image as a cue. We set out to simulate our hypothesis and assess whether it is a computationally efficient strategy for this task. To this end, we leverage existing artificial Spiking Neural Networks (SNNs) trained with \textit{Spike Timing Dependent Plasticity} (STDP) learning and simulate a ``teacher'' that simultaneously shows to the SNN, during learning, the images and their corresponding names. Then during test, backpropagated action potentials are used to retrieve the right name. We compare the accuracy of a naming mechanism employing our hypothesis to that of a machine learning classifier. 
This section focuses on our modeling. We first describe recent existing SNN models we build on (Sec.~\ref{background:SNN}). We then critically review their limits and plausibility (Sec.~\ref{plausibility}). Finally, we detail how we use them to build our model (Sec.~\ref{our_model}).

\subsection{Image classification with existing Spiking Neural Networks}\label{background:SNN}
SNNs are a class of biologically inspired computational models in which spiking neurons communicate information through individual spikes that propagate from one neuron to the next. Such spikes simulate APs, happening when the membrane potential of the neuron crosses a certain threshold. In reality, both the rates at which spikes are generated and the temporal patterns of spikes are believed to carry information about the input stimuli~\cite{tavanaei2018deep,gerstner2002spiking}. The artificial SNNs we leverage in this paper simulate a simpler version of this process, yet still offering higher biological plausibility~\cite{ghosh2009spiking} compared to other artificial models. Indeed, for training, the SNN we leverage uses the more plausible STDP learning rule
~\cite{taylor:stdp,markram:stdp,caporale2008spike,huang2014associative,mcmahon2012stimulus}. Under this rule, synaptic weights are updated according to the relative spike times of pre and post synaptic neurons: if the pre-synaptic spike occurs slightly before the post-synaptic spike, then a persistent strengthening of synapses called long-term potentiation (LTP) occurs~\cite{tavanaei2018deep}. In the other case, the result is a long-term depression (LTD) which leads to a persistent depotentiation of synapses.

Two recent models in particular provided background for our simulations~\cite{kheradpisheh2018stdp,mozafari2018first}. We build in particular on that of Kheradpisheh \textit{et al.}~\cite{kheradpisheh2018stdp} which achieved impressive accuracy on simple datasets. We reuse ``as-is'' its feature extraction layers. The latter are illustrated in the ``feature learning (STDP)'' upper part of Fig.~\ref{fig:model}.
As can be seen, it consists of consecutive layers of neural processing. The first is a temporal coding layer, that is meant to somewhat simulate retinal ganglion cells firing moments. It is followed by a cascade of convolutional and pooling layers to extract visual features. In more details, the first layer is responsible of encoding the input signal into discrete spike trains in the temporal domain. For this, it uses Difference of Gaussian (DoG) filters.\footnote{often used to grossly approximate the spatial visual processing in the retina.} This layer detects positive and negative contrasts in the input image and encodes them in spike latencies, according to their strengths. Next, each neuron in the convolutional layer receives input spikes from the neurons located in a certain window and emits a spike when its potential reaches a specific threshold. Pooling layers perform a nonlinear max pooling operation in which they only propagate the first spike emitted. 
In this model, STDP learning only occurs in convolutional layers and it is done layer by layer. For each image presented to the neural network, there is a ``competition'' between the neurons of a convolutional layer and those which fire earlier trigger STDP and learn the input pattern. Finally, the last layer is a global pooling layer which performs a global max pooling. The role of these feature extraction layers is just to learn visual features: they are trained without name labels by propagating many images through the layers and adjusting the weights with STDP.

Next, unlike what happens within our hypothesis, in Kheradpisheh's model~\cite{kheradpisheh2018stdp}, the trained output of this final layer is used to train a linear Support Vector Machine (SVM) classifier. The SVM classifier is of course not biologically plausible but the goal of Kheradpisheh et al. was only to assess the ability of SNNs and STDP to extract salient visual features that are good enough to discriminate images. And they actually found that they were good enough in terms of classification accuracy: their implementation reached 99\%, and 98.4\% of accuracy in the face/motorbike and MNIST datasets, respectively. We reproduced their results using Perez's available implementation~\cite{SDNN:implementation}. After some search of best parameters\cite{Our:SDNN:implementation}, we reached around \textit{96\%} of accuracy on the face/motorbike dataset with an SVM classifier.

Finally, worth mentioning, Mozafari et al.~\cite{mozafari2018first} proposed a 4 layers SNN with STDP, whose classification last layer is trained this time using reinforcement learning, instead of the SVM classifier. 
Their final layer is a decision making layer that performs a global pooling operation.  Each neuron in it is assigned to a category and the neuron which fires first indicates the network decision. This work, unlike the previous one does not thus rely on an external, biologically not plausible, classifier. Indeed, weight change in the last layer is modulated by a reward/punishment signal which depends on the correctness/incorrectness of the network's decision. However, the paper lacks plausibility in that it does not answer the challenging question of how and who generates the reward and punishment signals and more crucially: how does it ``know'' which neurons to punish and which neuron to ``reward''. Later, we present instead an end-to-end model, at the neuronal level, from learning associations to naming. Indeed, we postulate that the simple repeated co-occurence of signifier and signified is enough to tie them together as in passive learning, and this, in a bidirectional way (both naming and understanding). Hence in our framework, first, no unknown reward signal or mechanism is needed. Second, the naming of the object does not implicate a feed-forward mechanism but rather the backpropagation of action potentials. 

\subsection{Plausibility of the above SNN models}\label{plausibility}
Next, before using Kheradpisheh's model~\cite{kheradpisheh2018stdp} as a basis, we briefly discuss its (lack of) plausibility, as this allows us to later better gauge the plausibility of our hypothesis. 
In a nutshell, we are aware that the SNN model above lacks plausibility in many aspects, despite being unsupervised, and despite using a simple rule like STDP learning. For example, it uses only a spike-time neural coding. It also uses convolutional neural networks (CNNs) with weight sharing, which is biologically not plausible. However, all these ``problems'' do not impact our hypothesis since we are mainly interested in the last layers of the neural networks (where the retrograde signaling or backpropagation of the action potentials will actually initiate). Besides, and very interestingly, recent work~\cite{ott2020learning:sharing} has shown that training using ``properly translated data'' such as the case of ``correlated'' images in video relieves the need of using CNN weight sharing, and results in an approximate form of it.

Our position here is as follows. The above simple neural network model trained only with unsupervised STDP learning achieves good performance on what was 20 years ago a difficult problem. This means that it extracts visual features of fairly good quality. The latter are of course far from being perfect: the accuracy does not reach 100\% even on the simplest motor/face dataset. Nonetheless, we set out to verify if our hypothesis with STDP learning can successfully use these same features to find the right name association. For fairness, we compare our performance results to those of the SVM classifier.

\subsection{Our three-step model}\label{our_model}
\begin{figure}[t]
    \centering
    \includegraphics[width=\linewidth]{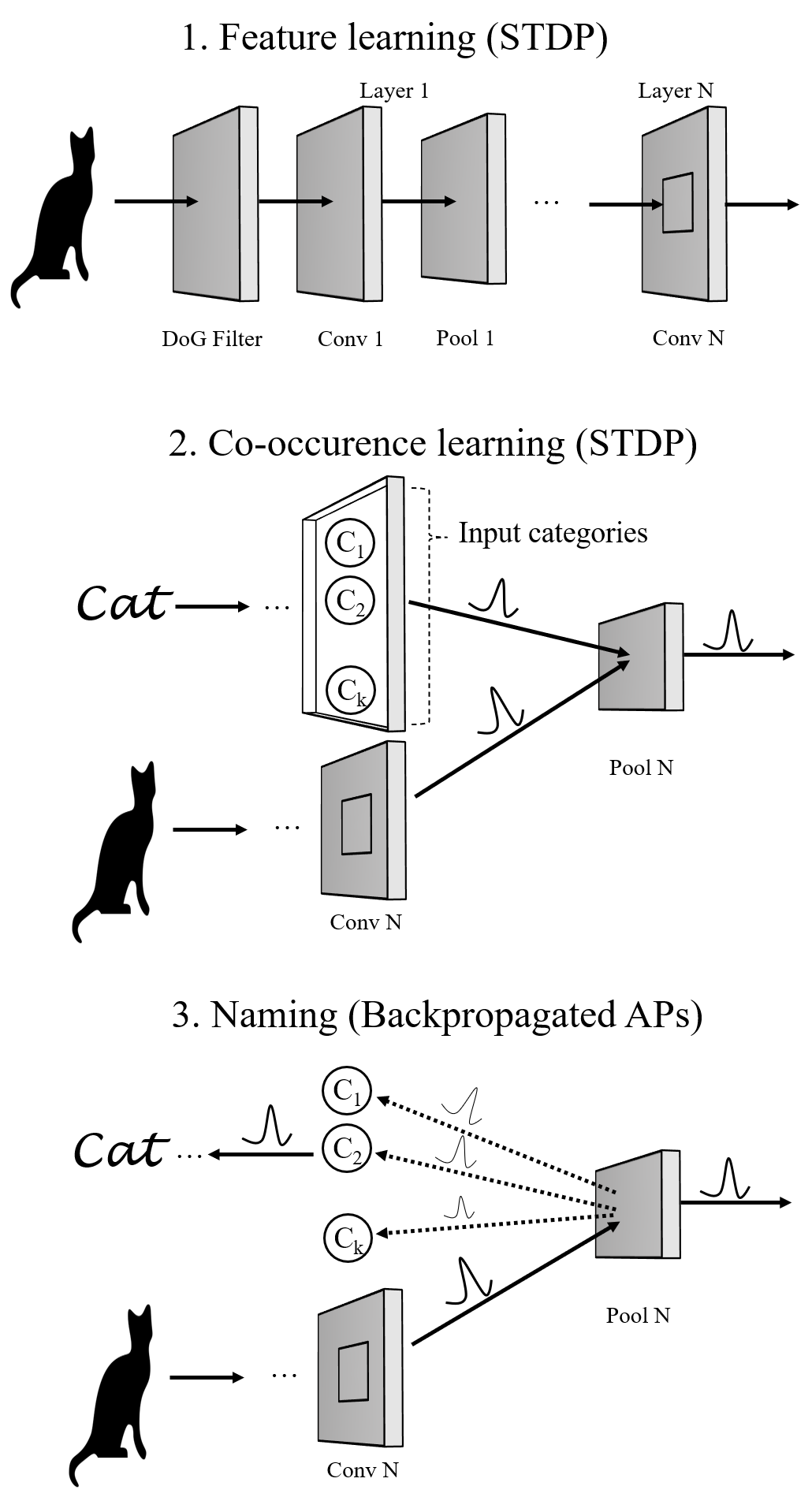}
    \caption{Learning under our hypothesis, simulated with the proposed SDNN with its three main parts.} 
    \label{fig:model}
\end{figure}

Under our hypothesis, successful name association comprises three steps, two for learning and one for recollection, as modeled in Fig.~\ref{fig:model}. 
The first \textit{feature learning} step is completely unsupervised and learns through repeated exposure to visual stimuli to extract salient features (e.g. lines, shapes etc) to discriminate visual content. In the brain, such learning is supposed to happen early in life. And if, for some reason or another, one is not exposed enough to visual stimuli, such learning does not happen, which leads to cortical blindness. 
As already mentioned, we reuse the SNN model described above~\cite{kheradpisheh2018stdp} to model it. 

The second step is a semi-supervised learning one, whereby a teacher shows a learner an image and the right name that refers to it. We call this \textit{co-occurence learning}. Humans for instance can learn from a single example to map a new object to its new name. Sometimes, when no external reinforcement happens, the exposition needs to be repeated multiple times until it is remembered. 
We model this step by simply adding a new ``categories layer'' and emulating the right spike each time an image is propagated through the SNN, i.e. generate a spike for ``cat'' category neuron while the cat image is propagated through the SNN.

The last step is simply the recollection of the name. It is in this step that a retrograde signaling from \textit{all neurons} in the backpropagation root layer is sent backwards to the categories layer. The neuron(s) which receive the highest ``vote'' signal the network decision. We now describe the three steps of Figure~\ref{fig:model} in more details.

\subsubsection{Feature Learning}
For feature learning, we used almost as-is the SNN model above~\cite{kheradpisheh2018stdp}. The reader can refer to the original paper for more details. This phase starts with the input image being encoded into discrete spike events in the temporal domain. Each ``spatial position'' or pixel in the input image will have its own spiking time. Such time-encoding process is performed by using Difference of Gaussian (DoG) filters, where spike times are computed according to the output of the DoG filter. More precisely, let $r$ be the value at a certain position index after having applied the DoG filter. Then, the firing time $t$ is defined to be $t = \frac{1}{r}$. This corresponds to encoding higher contrast areas of the image to lower spike times (i.e., latency is inversely proportional to the contrast). As a result, each single image is transformed into several waves of spikes that propagate, one by one, through the layers: spikes that signal higher contrast areas being the first to enter the network. 
Next, Convolutional layers are arranged in a feedforward manner.
Between two consecutive convolutional layers, a pooling layer performs a max operation to compress visual data and provide translation invariance. The task of a neuron in a pooling layer simply consists in propagating the first spike received from a receptive window of the previous convolutional layer.
Neurons in all the convolutional layers are \textit{non-leaky integrate and fire} neurons. They integrate input spikes and emit a spike as soon as they reach their threshold. The latter is a hyperparameter to set. 
Immediately after a spike occurs, weights are updated accordingly, using a simplified version of the STDP learning rule. Let $i,j$ be the indices of the post and pre-synaptic neurons, respectively, and let $t_i,t_j$ be their corresponding spike times. The synaptic weight $w_{ij}$ is updated by adding a modification factor $\Delta_{ij}$ computed as follows according to a simplified version of STDP~\cite{kheradpisheh2018stdp,masquelier2007unsupervised}.
\begin{equation}
\begin{aligned}
\Delta_{ij}=
\begin{cases}  \ \  \alpha^+ \cdot w_{ij}\cdot(1-w_{ij}),  \text{ if } t_j \leq t_i\\ 
- \alpha^- \cdot w_{ij}\cdot(1-w_{ij}), \text{ if } t_j > t_i
\end{cases}  \label{eq:stdp}
\end{aligned}
\end{equation}
$\alpha^+,\alpha^- \in \mathbb{R}_{\geq 0}$ are two parameters that specify the learning rate or by how much the weights are changed. The latter factor impacts a lot the learning. Indeed, small values would lead to a slow learning process, they simulate a neural network that is confident in its prior ``beliefs and decisions'' (weights). High values allow to learn very fast information about the current stimuli, but they can have as a consequence to ``forget'' what they learned with previous stimuli. 
Note that this simplified version of the STDP rule does not take into account the absolute time difference between post and pre-synaptic spikes. What matters instead is only the order, or the sign of the difference. In practice, this is not a problem for our model.

This feature learning process goes on, by propagating training images one by one. Each time the different spiking waves of a new image have been fully processed
, and weights updated and stored, the potential of each neuron is reset to 0, preparing for the next image.
Initially, the synaptic weights are chosen at random from a normal distribution with some mean and standard deviation\footnote{In practice, we try different initializations~\cite{Our:SDNN:implementation} and pick the best}. The STDP rule ensures that they always remain in the range $[0,1]$.
Within each image, learning is done layer by layer: learning at layer $\ell$ begins when learning at layer $\ell-1$ has terminated.

The intent of this feature learning phase is to learn the synaptic weights of each neuron in all the convolutional layers. As observed previously with this SNN model~\cite{kheradpisheh2018stdp}, neurons in the first layer converged to the simple four oriented edges. Neurons in the successive layers learned more complex ones by integrating spikes from previous layers. We remind that this phase is totally unsupervised as the network only learns frequent features associated with images and it requires no knowledge about the input image categories. The next learning step includes such categories, we qualify it as semi-supervised. 

\subsubsection{Co-occurrence learning} 
Once the SNN has learned the right weights and hence visual features, the second co-occurence learning step can begin. 
For this step, we assume as per our hypothesis, that there is a layer which encodes the object categories or names (equivalent of sparse neurons that respond to image or sound signifiers or both). The latter layer is connected, as shown in the figure, to the last layer of the image processing network. 
Then, an image (e.g. a cat as illustrated in the figure) is propagated through the image processing neural network, while at the same time, the neuron\footnote{we simulate a single neuron for simplicity, but same reasoning applies multiple ones} which represents its signifier or name is activated simultaneously. 

In more details, as in the previous phase, train images are considered one by one. Using the weights learned in the first phase, each image passes through the network until it reaches the last layer where a max pooling operation is performed. 
During co-occurence learning, a neuron of the last pooling layer would thus receive two spikes: one propagated by the neuron associated to the class of the image, and one input from the last convolutional layer. This entire process simulates the teacher that simultaneously shows the image and its name.

Note that, implementation-wise~\cite{Our:SDNN:implementation}, this is equivalent to having a matrix with the same shape as the last pooling layer that is associated to each image category (one weight per category neuron per neuron in the backpropagation layer).
As in the first phase, weights are initially random and are updated only using the STDP learning rule, as defined in \eqref{eq:stdp}. As a result of this STDP learning, according to the order of post and pre-synaptic spike times and to the index of the spiking neuron, what will happen is the following: the weight matrix of the correct image category is always strengthened (LTP), while the weight matrices of the other categories will be weakened (LTD).

Contrarily to the previous one, this phase is supervised in that it requires knowledge of the image category in order to be able to link it with the corresponding image features. Indeed, the aim here is to learn associations between the previously learned features and image categories by using only the simple STDP rule.
At the end of the second phase, training has completed and the SNN can proceed to the naming task using the backpropagation principle. 

One extreme version of this second phase is what is called one-shot learning: the network is given only a single example of each category. We will vary the number of such training examples in Sec.~\ref{experimental}, effectively trying one-shot and few-shot learning scenarios, hence why we consider this task to be ``semi-supervised''. 

\subsubsection{Naming}
Once the \textit{learning} is done with the previous two phases, we are ready now for the \textit{naming} task, following the principles of backpropagated action potentials. 

In this task, the image to name is first propagated through the fully trained neural network until spikes start to happen in the last pooling layer, which is our backpropagation root layer. We consider that all neurons in this layer are source pointer neurons. This means, as per our hypothesis, that we allow them (whenever they fire) to send backpropagated action potentials, modulated by the presynaptic weights learned in the previous second step, to the previous layer that encodes the labels or names. Neurons in the ``categories/signifiers" layer will integrate such received signals and the category which has the highest vote is the retained name for the image. Namely, let $C_i$ for $i=1,..,k$ be the class associated to the neuron with the highest class score. Then, $i$ is chosen to be the class the image belongs to. It is this score that can be used as the accumulated potential that brings the right neuron closer to its firing threshold, leading when it fires to the class decision\footnote{In practise, implementation-wise, the score is simply the sum of weights. Indeed, the SNN model considers a spike to be a binary decision that happens when the accumulated potential reaches a threshold (i.e. we do not consider the rate). We tried a version that uses the exact value of the internal potential instead of the unit value 1 but results were similar.}. 

We show later in the next section that such a simple mechanism allows to label the images as accurately as the SVM. More interestingly, by using high learning rates during the first co-occurence (e.g. neuromodulating to increase synaptic strength), it is possible to learn to name objects, with maximum accuracy, by showing the neural network only a single instance of the image class; an extreme learning task in which the SVM classifier seems to have more difficulty.

\begin{figure}[t]
    \centering
    \includegraphics[width=\linewidth]{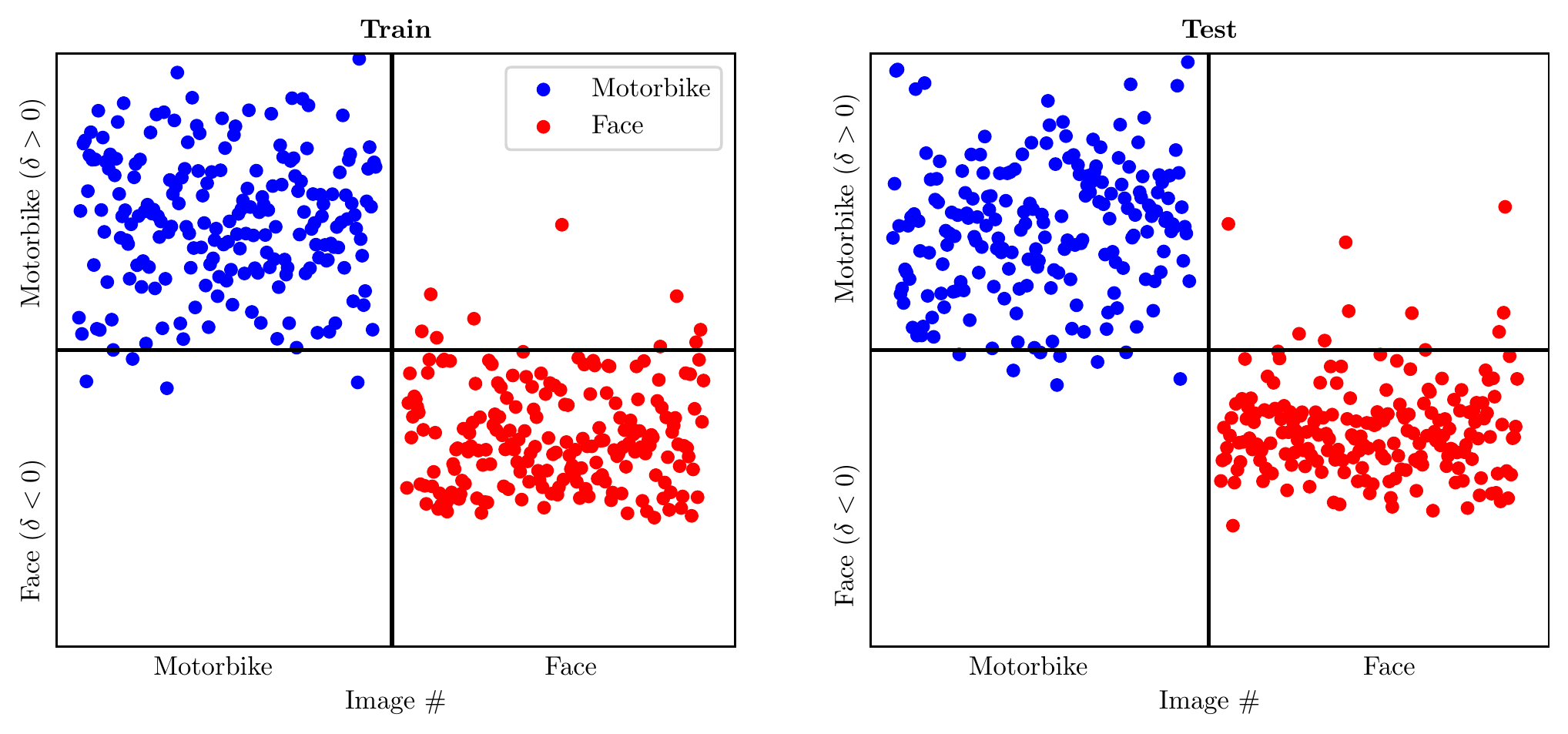}
    \caption{Difference between the class scores for each image in the Train (Left) and Test (Right) datasets.}
    \label{fig:res_scores}
\end{figure}
\section{Simulation results}\label{experimental}
We now evaluate the accuracy of the SNN model above when using our hypothesis to learn and name (the three steps described in Fig.~\ref{fig:model}). We compare its accuracy to that of Kheradpisheh \textit{et al.}~\cite{kheradpisheh2018stdp} (the same feature learning as step 1 above, followed by an SVM classifier on the feature vectors).

\subsection{Experimental Setup}
Experiments have been performed on a server with 5 Intel 2.10 GHz CPU, 32 GB of memory and a GPU Nvidia Tesla P100 SXM2 with 16 GB of dedicated memory. We evaluate the accuracy reached by our model using the Caltech motor face dataset~\cite{fei2007learning} considering two classes: Faces and Motorbikes.

\subsection{Overall accuracy with backpropagated APs}
We first focus on the overall accuracy and the case where we show the SNN many examples of each class during learning. In particular, for each class, we select 398 images among which 200 are reserved for training (feature learning and co-occurence learning likewise) and 198 are left for testing.
After a search of parameters~\cite{Our:SDNN:implementation}, we set $\alpha^+$ and $\alpha^-$ to $0.007$ and $0.003$, respectively. The firing thresholds for the first, second, and third convolutional layers are set to $6, 21$, and $10$. Similar to prior work~\cite{kheradpisheh2018stdp}, max pooling is not performed for this Caltech dataset because images have low resolutions. Our last layer uses thus a pooling window of size $1$x$1$. 
Finally, the synaptic weights of the class matrices are initialized at random from a normal distribution with mean $0.5$ and standard deviation $0.05$.

\textit{In this setting, using the backpropagation-based recollection, we reach an accuracy of $96.7$ and $95.7\%$ on the train and test datasets, respectively. This performance is on par with that of the SVM classifier ($96\%$), thus confirming the computational efficiency of backpropagated action potentials.}

In more details, Fig.~\ref{fig:res_scores} shows the class scores for both train and test datasets. Recall that these scores correspond to the ``votes'' received by backpropagation from the last layer. Each point in the figure corresponds to an image with motorbike points placed on the left and faces on the right. Values in the ordinate represent the difference between the scores associated to the Motorbike and Face classes. Therefore, images with positive values are associated to the Motorbike class, while images with negative values are associated to the Face one. 
As it can be observed, backpropgated action potentials allow to separate the two classes in a net way for most of the images. However, for some of the images, this distinction is not clear. We believe as we discussed earlier that the problem comes from the SNN feature extractor which, although performing, does not yet learn good representations. Nonetheless, only few images are classified incorrectly with a large relative error. Such a good performance shows overall the computational plausibility of the backpropagation-based recollection.

\subsection{Accuracy in few-shot learning}
We now focus on the case where the ``teacher'' shows the SNN only few examples.\\

\begin{figure}[t]
    \centering
    \includegraphics[width=0.7\linewidth]{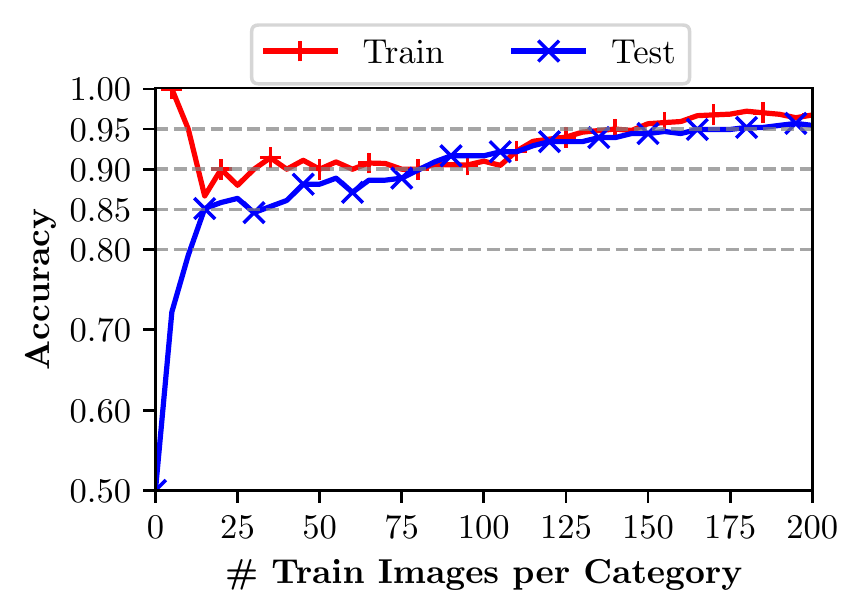}
    \caption{Accuracy as a function of the number of train images per category used in the co-occurrence Learning phase.}
    \label{fig:scores}
\end{figure}

\noindent\textbf{Varying the Number of Images in co-occurence learning}
We first vary the number of labeled examples given in the co-occurence based learning phase. In the remainder, whenever we talk about training, we refer to the supervised ``teacher-based'' co-occurence learning where a label is given.
Fig.\ref{fig:scores} shows the accuracy on the train and test datasets as a function of the number of labeled train images. We vary the number of labeled train images per experiment from $5$ to $200$. As the number of train images increases, the accuracies on both the train and test images increase as well. The train and test scores pass from $90\%$ and $85\%$ after 25 images per category up to the $96.75\%$ and $95.7\%$, from above, after having used the full train dataset.
This shows that the SNN learns but slowly as we feed images and labels. One way however to speed up this process is to increase the learning rates.

\begin{figure}[t]
    \centering
    \includegraphics[width=0.7\linewidth]{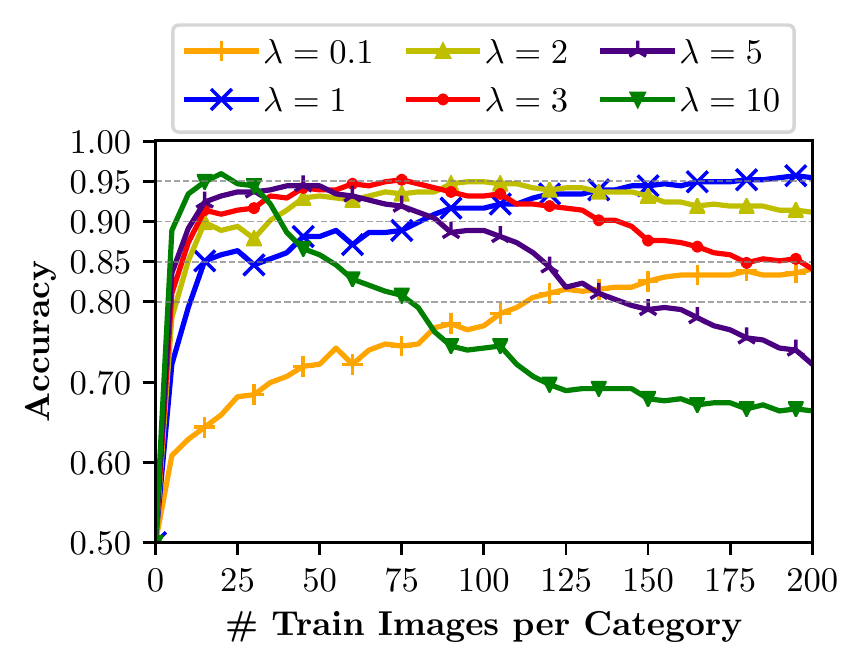}
    \caption{Accuracy on the test images as a function of the number of train images per category used in the co-occurrence Learning phase (increased number of shots in a few-shot learning task).}
    \label{fig:lambda}
\end{figure}

\mbox{}\\
\noindent\textbf{Varying the learning rate.}
Increasing the learning rate can simulate a neuromodulatory action that strengthens a connection suddenly, without the need for repeated exposure. We hence vary $\alpha^+$ and $\alpha^-$ used in the co-occurrence phase and observe a considerable impact on the accuracy. Fig. \ref{fig:lambda} illustrates the impact of the modulated learning rate on the accuracy level reached as a function of the number of train images used. As a baseline, we use $\alpha^+=0.007,\alpha^-=0.003$ and we multiply both values by some factor $\lambda \in \{\frac{1}{10},1,2,3,5,10\}$. 

When $\lambda < 1$, the learning is obviously slower. Indeed, the accuracy varies from $50\%$ with $0$ training images (i.e. random), to $85\%$ using the entire train dataset and grows in a linear way. It will probably reach higher values with more training time. When $\lambda = 1$, we reach the maximum\footnote{This is expected as the default learning rates are those that maximize the performance on the training set.} possible test score using all the train images.  

An interesting behaviour can be observed when $\lambda > 1$. The learning is at first faster, as it can reach high accuracy after having seen only a small sample of train images. However, it then starts decreasing as the number of train images increases.  We recall that the used STDP rule keeps the weights within the range $[0,1]$. Thus, starting with high values for $\alpha^+$ and $\alpha^-$ allows to faster associate discriminant features with the images categories. However, as the number of train images increases, weights might become less helpful if they tend to reach the maximum value of $1$ and to have thus smaller intermediate values. The scores would become closer and less distinguishable. Another factor is that, as we will see later, the SNN is better at recognizing and learning from certain particular images compared to others (see description of Fig.~\ref{fig:compare}). Hence, being exposed to a good image with a higher learning rate leads to reaching a good accuracy. But being exposed afterwards to a ``bad'' image will lead to unlearning the good weights, thus decreasing the performance.

When $\lambda=10$ and with only $20$ train images per category, it is possible to reach an accuracy of $95.5\%$. With $\lambda=2,3,\text{and\ } 5$, the number of necessary train images per category to reach the same accuracy are $45,75,\text{and\ } 100$, respectively. These results hint to the following direction: the ``best'' approach in terms of few-shot learning would be to first start with a high learning rate or $\lambda$ but then to stop changing the weights by either suddenly decreasing $\lambda$, or by simply freezing the learning and making the network always stick to the old beliefs.

\subsection{One-shot learning: Machine learning vs. Backproagation}
In this final section, we set the bar high and propose to train and test the SNN in a one shot learning task, meaning that we show the neural network only one single image from each class, together with their correct names. We then test the accuracy of the network on the entire 198 images of the test set. We compare SVM and our model on this task. 

As explored earlier, reaching good performance in this task goes through even higher learning rates than tried previously. We experiment with various $\lambda$s and various (motorbike,face) image couples. We find for instance that $\lambda=65$ yields good results among many other values, so we pick it. But this is where we interestingly discover that the performance highly depends on which couple of (motorbike,face) were used for the one-shot learning task. We found that, using the backpropagation-based recollection, certain single couples of motorbike and face yielded an accuracy of 96.2\% on \textit{all the remaining unseen} 198 test images. Note that this is higher than what we achieved earlier when training with all images and not only one or few shots (we obtained 95.7\% on test set). At the same time, the maximum we could achieve with the SVM classifier on a single example was 84.5\% of accuracy.

\begin{figure}[t]
    \centering
    \includegraphics[width=0.7\linewidth]{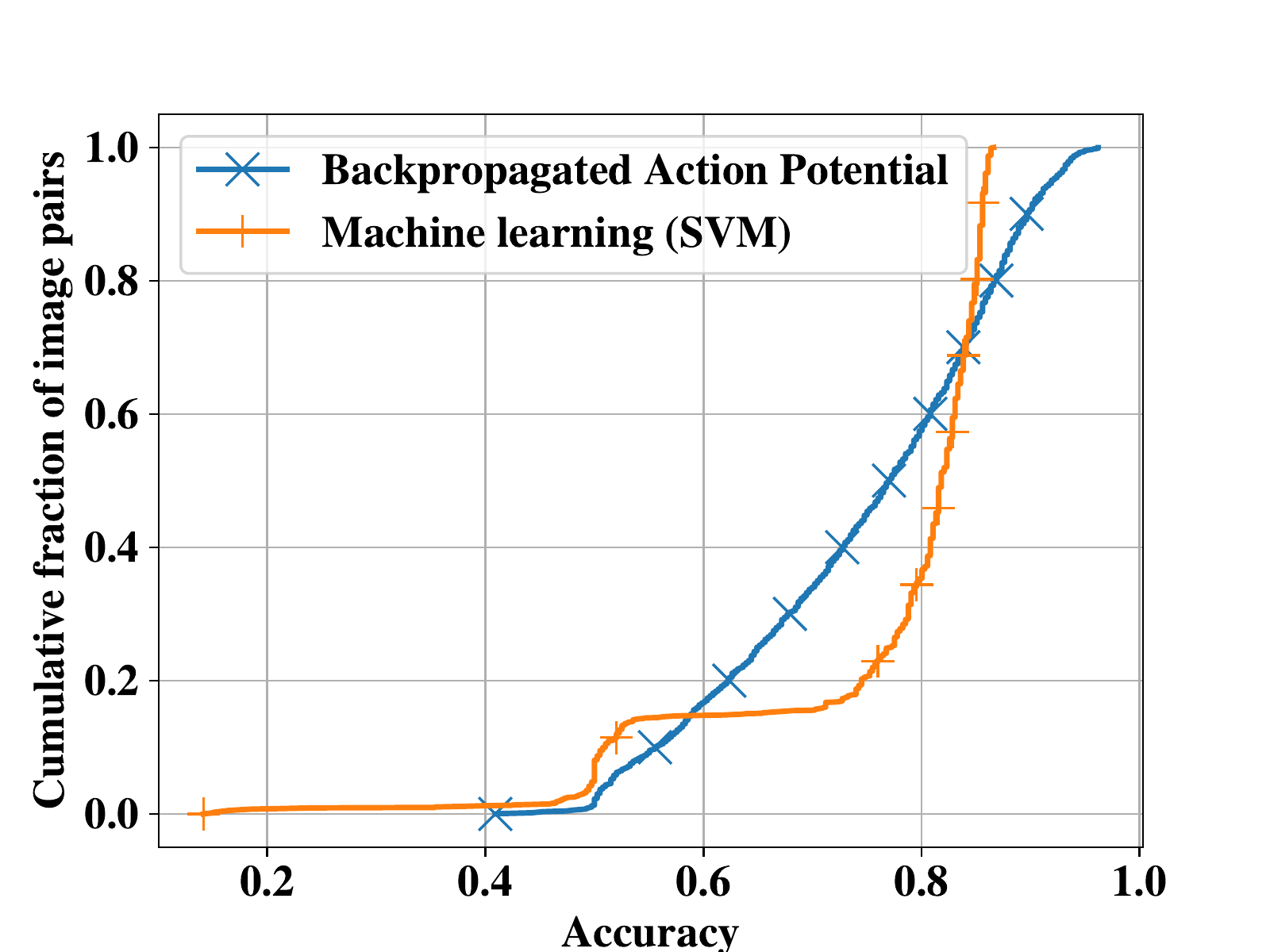}
    \caption{Comparison of ML and our hypothesis in one-shot learning tasks (CDF across more than 1500 experiments where each time, a single different pair of labeled (motor,face) images is shown during training). Evaluation is done on the usual test sets of around 400 images. When using the backpropagation-based hypothesis, there exists specific (motor,face) examples that, shown to the SNN only once, yield maximum accuracy (96\%) on the entire unseen test set.}
    \label{fig:compare}
\end{figure}

To assess this more systematically, we test both SVM and backpropagation in the one-shot exercise on more than 1500 different image couples of motorbike and face photos. We plot in Fig.~\ref{fig:compare} the resulting empirical cumulative distribution functions. The figure shows that both approaches yield different distributions, with most of SVM results being less dispersed around a little higher than 80\% accuracy. 
The observed curious disparity between pairs of images suggests that the SNN models are still not good enough in feature extraction. The learned representations are probably not as invariant as they should be, hence the variability and differences between images. Further investigating~\footnote{visual inspection of ``good'' and ``bad'' images did not uncover any peculiar character to distinguish them} the differences between these successful and less successful image couples might help enhancing current SNN models.

\textit{For these reasons, our simulations should be viewed only as an additional argument that shows the computational effectiveness of the backpropagation-based recollection mechanism, compared to an SVM classifier on the same extracted visual features. Our results above show indeed that with the right images used for training, backpropagation outperforms by far the SVM in the one shot task (96.2\% accuracy on 198 images against only 84.5\%).} 
\section{Discussion}\label{sec:discussion}
For more than a century~\cite{berlucchi1999some}, information processing in the brain has been mainly believed to follow the forward pre to post-synaptic neurons direction. In this work, we emitted the hypothesis that the backpropagation of action potentials in the reverse direction mediates all ``offline'' generative tasks where the simultaneous activation of specific targeted populations of neurons is needed. This is, we claimed, the case of the retrieval of past memories or mental images, retrieval of signification of words, retrieval of names and even the mixture of distinct past memories into imagination.
We reviewed in sec.~\ref{evidence} evidence that calls for taking the hypothesis seriously (see summary in Sec.~\ref{summary:arguments}). As an added bonus, we computationally showed in Sec.\ref{experimental} that our hypothesis can be more efficient than a machine learning algorithm in retrieving the category name of an object. If confirmed true, this would have tremendous implications, considerably improving our understanding of neural encoding and how high cognitive functions are implemented from a low-level neural perspective. 
\subsection{Possible implications}\label{sec:implications}
The first big implication of this hypothesis is the promise to bring answers to the neural encoding problem, closing the old debate in cognitive sciences between localist and distributed representation theories. If our hypothesis is true, the answer to the representation problem becomes simpler: (i) representations of concepts are distributed over a population of neurons but at the same time, (ii) there exists highly selective neurons that respond to unique concepts, but that serve a specific retrieval purpose. The latter act as hubs between various related concepts, playing the role of source pointer neurons to retrieve the entire concept's features (encoded by an entire population of neurons). For example, there should exist relatively few neurons that selectively respond to the image of a cat. But such neurons serve as pointers or hubs to easily connect such image to other related memories. Similar sparse neurons act for instance as pointers to retrieve the visual features of a cat through backpropagated APs that travel backwards to reactivate selectively the neurons that represent the right lines, shapes\footnote{Such lower level features are eventually shared by other objects and do not respond only to the presence of a cat.} and colours that define a cat. Hence, to be able to recall through mental visual imagery the image of a cat, the brain does not need to activate only the sparse neurons that respond to ``cats'', but, as the optogenetic studies above also found, the activation of an entire larger population is needed. And if neurons that represent, say, vertical lines, are not activated during this process, the recalled mental image will lack them. 

Now beyond vision, the cat's toy example could apply to any mental state and any couple of \textit{(lived  stimuli, later memory of that stimuli)}, let them be smells, affects or even impressions of movements. In accordance with the principle of grounded cognition~\cite{barsalou2008grounded}, for which we believe our hypothesis framework applies, discrete concepts are grounded in the sensorimotor experiences that were encoded with them, such that the activation of a signifier of a concept leads to the activation of the experiences that are grounded with it. Hence, the cue that is the words ``moving" or ``tickling" is correlated with areas that encode actual moving or tickling.

Next, another related hard problem in cognitive sciences that can benefit from our hypothesis is the binding problem or how does the brain binds higher-level concepts to more elementary ones and particularly, how does it associate the right features (e.g. colors) to the right discrete objects or concepts, for example in an image composed of many objects. The fact that the brain needs some time to correctly perform the binding~\cite{von1999and:binding} suggests that this operation is not forward-based but generatively and iteratively happens later in a second stage. If our hypothesis is correct, this should happen through slow and repetitive runs of top down action potential backpropagation, starting from the right source pointer neurons that define uniquely the discrete object, all the way backwards, activating all the neurons that describe its attributes.
In fact, two famous competing (and high level) theories on this problem are still the feature-integration theory~\cite{treisman1980feature} and temporal synchronization theory~\cite{milner1974model,shadlen1999synchrony}. Our hypothesis offers perspectives to reconcile them as well. Indeed, both admit the involvement of different runs of bottom up and top down hierarchies (e.g. attention in feature integration theory) to implement the binding problem. However, the exact physical mechanisms that would implement them are still unknown today. 
Our hypothesis constitutes a good candidate for such mechanisms, with backpropagated APs implementing such top down hierarchies to bind objects to their features. Moreover, under this new realm, one does not have anymore to chose exclusively between binding-by-synchrony and feature-integration theory: \textit{attention} with backpropagated APs could \textit{synchronously} activate selectively all the features related to a given discrete object. An implicit consequence of this is that attention itself is likely implemented through top-down backpropagated APs.
 
In general, backpropagation could constitute an ubiquitous and easy-to-implement simple mechanism that underlies a diverse set of cognitive tasks such as offline thinking or mind wandering, imagination, episodic memory retrieval and future episodic thinking. In our framework, imagination would be ``simply'' the resulting activation patterns of a mixture of usually unrelated concepts: for example, an imagined ``laughing cat'' results from simultaneously back-activating the ``laughing'' and ''cat'' concept neurons. The same applies for mind wandering, where backpropagated APs should induce activation patterns that follow ``the most likely neural pathways'', thus generating what seems to be coherent thoughts. 

Our hypothesis assumes the existence of (backpropagation root) areas where the flow reverses during cue-based retrieval (from the cue recognition networks, back to the networks of the memory trace). Our review of the literature confirms that a similar reversal happens in the MTL and the hippocampal formation. However, all this predicts also the existence of generators, or specialized centers, that release neuromodulators remotely to control the generation process by either inhibiting or facilitating backpropagation at such root layers. This is where more advanced modeling and computer simulations could help tremendously in future work. For example, it would be helpful to understand the interplay between backpropagation and forward propagation, since the first can result as well in feedforward propagation, that in turn might cause backpropagation etc. Such ``lateral'' activation patterns could be useful to find associated concepts, as opposed to ``digging into the details'' of a single concept.

Moreover, if proven true, our hypothesis would open new ways to better understand some recollection-related pathologies. In this case, understanding what factors impact the inhibition or excitation of backpropagated APs could help understanding probably-related disorders such as obsessive thinking or intrusive thoughts. Similar (rarer) dysfunctions happen in mental visual imagery as well, such as the absence (aphantasia) or excess (hyperphantasia) of visual imagery experiences~\cite{pearson2019human}. 
In all these cases, the neuromodulation mechanisms that control backpropagation should be first suspected. 

Last and not least, another implication is that the retrieval process is stochastic in nature: the retrieved memory trace looks like the original first perception but, depending on past experiences (and hence the weights of the neural connections due to past experiences), the reactivation might not be exactly the same. This can be seen most in the case of language where the same word (e.g. ``a screen'') was seen multiple times during encoding, in the presence of multiple similar stimuli (e.g. many different types of screens), and where the same concept is further grounded in different neural network connections from one subject to the other. 
Such considerations offer new perspectives to apprehend language as a system of common cues, useful to make others live intended target experiences through the invocation of the right cues.


\subsection{Verifiability and further investigations}
But before drawing a bright future to our hypothesis, at least two directions must be seriously taken to further investigate it and confirm its plausibility or infirm it. 

\subsubsection{Empirical methods}
This is perhaps the most important direction. We verified in the literature the existence of retrograde signals that satisfy many of the assumptions of our hypothesis. Further targeted empirical studies must verify the remaining ones so as to support or invalidate our hypothesis. In particular, first, it is crucial to understand if the backpropagating signal is stronger on paths with ``higher synaptic weights''. Second, it is necessary to measure how far-reaching the backpropagating signal can be, beyond solely the previous connection. In our literature review, \textit{we found no evidence for nor against the latter points}.

\subsubsection{Computational effectiveness}
Another line of work could verify in parallel the computational effectiveness of our hypothesis in implementing its other target goals. Here, we verified (as a first step) the ability of retrograde APs to perform the object recognition or the naming task.
We opted for this comparison because of the presence of a baseline to compare to (an existing image classifier) and a metric (accuracy) to quantify the computational power of the mechanism. 
Not simulated in this work, retrograde APs can be used symmetrically for the task of ``understanding": i.e. an activation of the signifier (word) neuron that leads automatically to the activation of the signified concept and its, say, visual features. This needs however, to rely on Artificial Neural networks that have good and plausible feature extraction capabilities. The SNNs we use are promising and close but they do not yet fully satisfy the last property (see Sec.~\ref{plausibility}). 

Finally, in the event that such artificial backward reconstruction works successfully with some SNN, our hypothesis could become then computationally verifiable for the imagination task. One interesting experiment could be to train artificial neural networks to recognize two separate concepts from images, exactly as we do for "Motor" and "Face" in Sec.~\ref{experimental}. Then, instead of backward constructing only one concept at a time, it would be interesting to simultaneously activate two concepts and see the effect on the backward reconstructed images. As exemplified above, one could activate a concept like "laughing" and a concept like "cat" and visualize the results of the ``competition" between backpropagating signals on the backward-constructed images. Note that similar work can be achieved with today's state of the art deep neural networks such as GPT-3's DALL·E~\cite{Dalle:gpt3} which uses a transformer decoder architecture. However, the latter employs supervised mechanisms that lack biological plausibility~\cite{grossberg1987competitive:backprop,crick1989:backprop,whittington2019theories}, which is our focus in this work.

\section*{Acknowledgement}
The author would like to acknowledge the contributions of Dr. Andrea Tomassilli for his help in the setup of the experiments and analysis of the results during his internship in fall 2018. We are grateful to Dr. Alessandro Finamore for interesting feedback on an earlier draft of the paper.



\bibliographystyle{ieeetr}
\bibliography{references}

\begin{thebibliography}{100}

\bibitem{saussure}
F.~De~Saussure, {\em Course in general linguistics}.
\newblock New York: McGraw-Hill., 1959.

\bibitem{quiroga2005invariant:anniston}
R.~Q. Quiroga, L.~Reddy, G.~Kreiman, C.~Koch, and I.~Fried, ``Invariant visual
  representation by single neurons in the human brain,'' {\em Nature},
  vol.~435, no.~7045, pp.~1102--1107, 2005.

\bibitem{connor2005friends}
C.~E. Connor, ``Friends and grandmothers,'' {\em Nature}, vol.~435, no.~7045,
  pp.~1036--1037, 2005.

\bibitem{pearson2019human}
J.~Pearson, ``The human imagination: the cognitive neuroscience of visual
  mental imagery,'' {\em Nature Reviews Neuroscience}, vol.~20, no.~10,
  pp.~624--634, 2019.

\bibitem{dijkstra2017distinct}
N.~Dijkstra, P.~Zeidman, S.~Ondobaka, M.~A. van Gerven, and K.~Friston,
  ``Distinct top-down and bottom-up brain connectivity during visual perception
  and imagery,'' {\em Scientific reports}, vol.~7, no.~1, pp.~1--9, 2017.

\bibitem{dentico2014reversal}
D.~Dentico, B.~L. Cheung, J.-Y. Chang, J.~Guokas, M.~Boly, G.~Tononi, and
  B.~Van~Veen, ``Reversal of cortical information flow during visual imagery as
  compared to visual perception,'' {\em Neuroimage}, vol.~100, pp.~237--243,
  2014.

\bibitem{VAE:1}
D.~P. Kingma and M.~Welling, ``Auto-encoding variational bayes,'' in {\em 2nd
  International Conference on Learning Representations, {ICLR} 2014, Banff, AB,
  Canada, April 14-16, 2014, Conference Track Proceedings}, 2014.

\bibitem{GANs}
I.~Goodfellow, J.~Pouget-Abadie, M.~Mirza, B.~Xu, D.~Warde-Farley, S.~Ozair,
  A.~Courville, and Y.~Bengio, ``Generative adversarial nets,'' in {\em
  Advances in neural information processing systems}, pp.~2672--2680, 2014.

\bibitem{grossberg1987competitive:backprop}
S.~Grossberg, ``Competitive learning: From interactive activation to adaptive
  resonance,'' {\em Cognitive science}, vol.~11, no.~1, pp.~23--63, 1987.

\bibitem{crick1989:backprop}
F.~Crick, ``The recent excitement about neural networks,'' {\em Nature},
  vol.~337, no.~6203, pp.~129--132, 1989.

\bibitem{whittington2019theories}
J.~C. Whittington and R.~Bogacz, ``Theories of error back-propagation in the
  brain,'' {\em Trends in cognitive sciences}, 2019.

\bibitem{kheradpisheh2018stdp}
S.~R. Kheradpisheh, M.~Ganjtabesh, S.~J. Thorpe, and T.~Masquelier,
  ``Stdp-based spiking deep convolutional neural networks for object
  recognition,'' {\em Neural Networks}, vol.~99, pp.~56--67, 2018.

\bibitem{mozafari2018first}
M.~Mozafari, S.~R. Kheradpisheh, T.~Masquelier, A.~Nowzari-Dalini, and
  M.~Ganjtabesh, ``First-spike-based visual categorization using
  reward-modulated stdp,'' {\em IEEE Transactions on Neural Networks and
  Learning Systems}, 2018.

\bibitem{SDNN:implementation}
N.~Perez-Nieves, ``Sdnn python.'' \url{https://github.com/npvoid/SDNN\_python}.
\newblock Accessed: 2020-11-08.

\bibitem{Our:SDNN:implementation}
``Backpropagation-based recollection hypothesis code.''
  \url{https://github.com/bendiogene/recollection_hypothesis}.
\newblock Accessed: 2021-01-10.

\bibitem{markram:stdp}
H.~Markram, J.~L{\"u}bke, M.~Frotscher, and B.~Sakmann, ``Regulation of
  synaptic efficacy by coincidence of postsynaptic aps and epsps,'' {\em
  Science}, vol.~275, no.~5297, pp.~213--215, 1997.

\bibitem{kruger2012deep:primate}
N.~Kruger, P.~Janssen, S.~Kalkan, M.~Lappe, A.~Leonardis, J.~Piater, A.~J.
  Rodriguez-Sanchez, and L.~Wiskott, ``Deep hierarchies in the primate visual
  cortex: What can we learn for computer vision?,'' {\em IEEE transactions on
  pattern analysis and machine intelligence}, vol.~35, no.~8, pp.~1847--1871,
  2012.

\bibitem{tulving1973encoding}
E.~Tulving and D.~M. Thomson, ``Encoding specificity and retrieval processes in
  episodic memory.,'' {\em Psychological review}, vol.~80, no.~5, p.~352, 1973.

\bibitem{bowers2009biological}
J.~S. Bowers, ``On the biological plausibility of grandmother cells:
  implications for neural network theories in psychology and neuroscience.,''
  {\em Psychological review}, vol.~116, no.~1, p.~220, 2009.

\bibitem{grandmother:psycho:neuro}
J.~S. Bowers, ``Grandmother cells and localist representations: a review of
  current thinking,'' {\em Language, Cognition and Neuroscience}, vol.~32,
  no.~3, pp.~257--273, 2017.

\bibitem{patterson2007you}
K.~Patterson, P.~J. Nestor, and T.~T. Rogers, ``Where do you know what you
  know? the representation of semantic knowledge in the human brain,'' {\em
  Nature reviews neuroscience}, vol.~8, no.~12, pp.~976--987, 2007.

\bibitem{svoboda1997vivo:anesthesized}
K.~Svoboda, W.~Denk, D.~Kleinfeld, and D.~W. Tank, ``In vivo dendritic calcium
  dynamics in neocortical pyramidal neurons,'' {\em Nature}, vol.~385,
  no.~6612, pp.~161--165, 1997.

\bibitem{bereshpolova2007:awake}
Y.~Bereshpolova, Y.~Amitai, A.~G. Gusev, C.~R. Stoelzel, and H.~A. Swadlow,
  ``Dendritic backpropagation and the state of the awake neocortex,'' {\em
  Journal of Neuroscience}, vol.~27, no.~35, pp.~9392--9399, 2007.

\bibitem{1998somadendritic:awake}
G.~Buzsaki and A.~Kandel, ``Somadendritic backpropagation of action potentials
  in cortical pyramidal cells of the awake rat,'' {\em Journal of
  neurophysiology}, vol.~79, no.~3, pp.~1587--1591, 1998.

\bibitem{stuart1997action}
G.~Stuart, N.~Spruston, B.~Sakmann, and M.~H{\"a}usser, ``Action potential
  initiation and backpropagation in neurons of the mammalian cns,'' {\em Trends
  in neurosciences}, vol.~20, no.~3, pp.~125--131, 1997.

\bibitem{vetter2001propagation}
P.~Vetter, A.~Roth, and M.~Hausser, ``Propagation of action potentials in
  dendrites depends on dendritic morphology,'' {\em Journal of
  neurophysiology}, vol.~85, no.~2, pp.~926--937, 2001.

\bibitem{waters2005backpropagating}
J.~Waters, A.~Schaefer, and B.~Sakmann, ``Backpropagating action potentials in
  neurones: measurement, mechanisms and potential functions,'' {\em Progress in
  biophysics and molecular biology}, vol.~87, no.~1, pp.~145--170, 2005.

\bibitem{williams2000action:TC}
S.~R. Williams and G.~J. Stuart, ``Action potential backpropagation and
  somato-dendritic distribution of ion channels in thalamocortical neurons,''
  {\em Journal of Neuroscience}, vol.~20, no.~4, pp.~1307--1317, 2000.

\bibitem{williams2000backpropagation}
S.~R. Williams and G.~J. Stuart, ``Backpropagation of physiological spike
  trains in neocortical pyramidal neurons: implications for temporal coding in
  dendrites,'' {\em Journal of Neuroscience}, vol.~20, no.~22, pp.~8238--8246,
  2000.

\bibitem{tao2001retrograde}
H.~W. Tao and M.-m. Poo, ``Retrograde signaling at central synapses,'' {\em
  Proceedings of the National Academy of Sciences}, vol.~98, no.~20,
  pp.~11009--11015, 2001.

\bibitem{tsubokawa1997muscarinic:neuromod}
H.~Tsubokawa and W.~N. Ross, ``Muscarinic modulation of spike backpropagation
  in the apical dendrites of hippocampal ca1 pyramidal neurons,'' {\em Journal
  of Neuroscience}, vol.~17, no.~15, pp.~5782--5791, 1997.

\bibitem{hoffman1999neuromodulation}
D.~A. Hoffman and D.~Johnston, ``Neuromodulation of dendritic action
  potentials,'' {\em Journal of neurophysiology}, vol.~81, no.~1, pp.~408--411,
  1999.

\bibitem{tulving1966availability}
E.~Tulving and Z.~Pearlstone, ``Availability versus accessibility of
  information in memory for words,'' {\em Journal of Verbal Learning and Verbal
  Behavior}, vol.~5, no.~4, pp.~381--391, 1966.

\bibitem{tulving1972episodic}
E.~Tulving {\em et~al.}, ``Episodic and semantic memory,'' {\em Organization of
  memory}, vol.~1, pp.~381--403, 1972.

\bibitem{frankland2019neurobiological}
P.~W. Frankland, S.~A. Josselyn, and S.~K{\"o}hler, ``The neurobiological
  foundation of memory retrieval,'' {\em Nature neuroscience}, vol.~22, no.~10,
  pp.~1576--1585, 2019.

\bibitem{tulving1983ecphoric}
E.~Tulving, ``Ecphoric processes in episodic memory,'' {\em Philosophical
  Transactions of the Royal Society of London. B, Biological Sciences},
  vol.~302, no.~1110, pp.~361--371, 1983.

\bibitem{schacter1978richard}
D.~L. Schacter, J.~E. Eich, and E.~Tulving, ``Richard semon's theory of
  memory,'' {\em Journal of Verbal Learning and Verbal Behavior}, vol.~17,
  no.~6, pp.~721--743, 1978.

\bibitem{nairne2002myth}
J.~S. Nairne, ``The myth of the encoding-retrieval match,'' {\em Memory},
  vol.~10, no.~5-6, pp.~389--395, 2002.

\bibitem{poirier2012memory}
M.~Poirier, J.~S. Nairne, C.~Morin, F.~G. Zimmermann, K.~Koutmeridou, and
  J.~Fowler, ``Memory as discrimination: A challenge to the encoding--retrieval
  match principle.,'' {\em Journal of Experimental Psychology: Learning,
  Memory, and Cognition}, vol.~38, no.~1, p.~16, 2012.

\bibitem{goh2012testing}
W.~D. Goh and S.~H. Lu, ``Testing the myth of the encoding--retrieval match,''
  {\em Memory \& cognition}, vol.~40, no.~1, pp.~28--39, 2012.

\bibitem{tanaka2014cortical}
K.~Z. Tanaka, A.~Pevzner, A.~B. Hamidi, Y.~Nakazawa, J.~Graham, and B.~J.
  Wiltgen, ``Cortical representations are reinstated by the hippocampus during
  memory retrieval,'' {\em Neuron}, vol.~84, no.~2, pp.~347--354, 2014.

\bibitem{liu2012optogenetic}
X.~Liu, S.~Ramirez, P.~T. Pang, C.~B. Puryear, A.~Govindarajan, K.~Deisseroth,
  and S.~Tonegawa, ``Optogenetic stimulation of a hippocampal engram activates
  fear memory recall,'' {\em Nature}, vol.~484, no.~7394, pp.~381--385, 2012.

\bibitem{vetere2019memory}
G.~Vetere, L.~M. Tran, S.~Moberg, P.~E. Steadman, L.~Restivo, F.~G. Morrison,
  K.~J. Ressler, S.~A. Josselyn, and P.~W. Frankland, ``Memory formation in the
  absence of experience,'' {\em Nature neuroscience}, vol.~22, no.~6,
  pp.~933--940, 2019.

\bibitem{smith2001environmental}
S.~M. Smith and E.~Vela, ``Environmental context-dependent memory: A review and
  meta-analysis,'' {\em Psychonomic bulletin \& review}, vol.~8, no.~2,
  pp.~203--220, 2001.

\bibitem{eich1995mood}
E.~Eich, ``Mood as a mediator of place dependent memory.,'' {\em Journal of
  Experimental Psychology: General}, vol.~124, no.~3, p.~293, 1995.

\bibitem{denny2014hippocampal}
C.~A. Denny, M.~A. Kheirbek, E.~L. Alba, K.~F. Tanaka, R.~A. Brachman, K.~B.
  Laughman, N.~K. Tomm, G.~F. Turi, A.~Losonczy, and R.~Hen, ``Hippocampal
  memory traces are differentially modulated by experience, time, and adult
  neurogenesis,'' {\em Neuron}, vol.~83, no.~1, pp.~189--201, 2014.

\bibitem{reijmers2007localization}
L.~G. Reijmers, B.~L. Perkins, N.~Matsuo, and M.~Mayford, ``Localization of a
  stable neural correlate of associative memory,'' {\em Science}, vol.~317,
  no.~5842, pp.~1230--1233, 2007.

\bibitem{sorensen2016robust}
A.~T. S{\o}rensen, Y.~A. Cooper, M.~V. Baratta, F.-J. Weng, Y.~Zhang,
  K.~Ramamoorthi, R.~Fropf, E.~LaVerriere, J.~Xue, A.~Young, {\em et~al.}, ``A
  robust activity marking system for exploring active neuronal ensembles,''
  {\em Elife}, vol.~5, p.~e13918, 2016.

\bibitem{lacagnina2019distinct}
A.~F. Lacagnina, E.~T. Brockway, C.~R. Crovetti, F.~Shue, M.~J. McCarty, K.~P.
  Sattler, S.~C. Lim, S.~L. Santos, C.~A. Denny, and M.~R. Drew, ``Distinct
  hippocampal engrams control extinction and relapse of fear memory,'' {\em
  Nature neuroscience}, vol.~22, no.~5, pp.~753--761, 2019.

\bibitem{khalaf2018reactivation}
O.~Khalaf, S.~Resch, L.~Dixsaut, V.~Gorden, L.~Glauser, and J.~Gr{\"a}ff,
  ``Reactivation of recall-induced neurons contributes to remote fear memory
  attenuation,'' {\em Science}, vol.~360, no.~6394, pp.~1239--1242, 2018.

\bibitem{ramirez2013creating}
S.~Ramirez, X.~Liu, P.-A. Lin, J.~Suh, M.~Pignatelli, R.~L. Redondo, T.~J.
  Ryan, and S.~Tonegawa, ``Creating a false memory in the hippocampus,'' {\em
  Science}, vol.~341, no.~6144, pp.~387--391, 2013.

\bibitem{tayler2013reactivation}
K.~K. Tayler, K.~Z. Tanaka, L.~G. Reijmers, and B.~J. Wiltgen, ``Reactivation
  of neural ensembles during the retrieval of recent and remote memory,'' {\em
  Current Biology}, vol.~23, no.~2, pp.~99--106, 2013.

\bibitem{guskjolen2018recovery}
A.~Guskjolen, J.~W. Kenney, J.~de~la Parra, B.-r.~A. Yeung, S.~A. Josselyn, and
  P.~W. Frankland, ``Recovery of “lost” infant memories in mice,'' {\em
  Current Biology}, vol.~28, no.~14, pp.~2283--2290, 2018.

\bibitem{waldhauser2016episodic}
G.~T. Waldhauser, V.~Braun, and S.~Hanslmayr, ``Episodic memory retrieval
  functionally relies on very rapid reactivation of sensory information,'' {\em
  Journal of Neuroscience}, vol.~36, no.~1, pp.~251--260, 2016.

\bibitem{jafarpour2014replay}
A.~Jafarpour, L.~Fuentemilla, A.~J. Horner, W.~Penny, and E.~Duzel, ``Replay of
  very early encoding representations during recollection,'' {\em Journal of
  Neuroscience}, vol.~34, no.~1, pp.~242--248, 2014.

\bibitem{johnson2009recollection}
J.~D. Johnson, S.~G. McDuff, M.~D. Rugg, and K.~A. Norman, ``Recollection,
  familiarity, and cortical reinstatement: a multivoxel pattern analysis,''
  {\em Neuron}, vol.~63, no.~5, pp.~697--708, 2009.

\bibitem{manning2011oscillatory}
J.~R. Manning, S.~M. Polyn, G.~H. Baltuch, B.~Litt, and M.~J. Kahana,
  ``Oscillatory patterns in temporal lobe reveal context reinstatement during
  memory search,'' {\em Proceedings of the National Academy of Sciences},
  vol.~108, no.~31, pp.~12893--12897, 2011.

\bibitem{ritchey2013neural}
M.~Ritchey, E.~A. Wing, K.~S. LaBar, and R.~Cabeza, ``Neural similarity between
  encoding and retrieval is related to memory via hippocampal interactions,''
  {\em Cerebral cortex}, vol.~23, no.~12, pp.~2818--2828, 2013.

\bibitem{staresina2012episodic}
B.~P. Staresina, R.~N. Henson, N.~Kriegeskorte, and A.~Alink, ``Episodic
  reinstatement in the medial temporal lobe,'' {\em Journal of Neuroscience},
  vol.~32, no.~50, pp.~18150--18156, 2012.

\bibitem{yaffe2014reinstatement}
R.~B. Yaffe, M.~S. Kerr, S.~Damera, S.~V. Sarma, S.~K. Inati, and K.~A.
  Zaghloul, ``Reinstatement of distributed cortical oscillations occurs with
  precise spatiotemporal dynamics during successful memory retrieval,'' {\em
  Proceedings of the National Academy of Sciences}, vol.~111, no.~52,
  pp.~18727--18732, 2014.

\bibitem{fulford2018neural}
J.~Fulford, F.~Milton, D.~Salas, A.~Smith, A.~Simler, C.~Winlove, and A.~Zeman,
  ``The neural correlates of visual imagery vividness--an fmri study and
  literature review,'' {\em Cortex}, vol.~105, pp.~26--40, 2018.

\bibitem{dijkstra2017vividness}
N.~Dijkstra, S.~E. Bosch, and M.~A. van Gerven, ``Vividness of visual imagery
  depends on the neural overlap with perception in visual areas,'' {\em Journal
  of Neuroscience}, vol.~37, no.~5, pp.~1367--1373, 2017.

\bibitem{dijkstra2019shared}
N.~Dijkstra, S.~E. Bosch, and M.~A. van Gerven, ``Shared neural mechanisms of
  visual perception and imagery,'' {\em Trends in cognitive sciences}, 2019.

\bibitem{dijkstra2018differential}
N.~Dijkstra, P.~Mostert, F.~P. de~Lange, S.~Bosch, and M.~A. van Gerven,
  ``Differential temporal dynamics during visual imagery and perception,'' {\em
  Elife}, vol.~7, p.~e33904, 2018.

\bibitem{st2015distributed}
M.~St-Laurent, H.~Abdi, and B.~R. Buchsbaum, ``Distributed patterns of
  reactivation predict vividness of recollection,'' {\em Journal of Cognitive
  Neuroscience}, vol.~27, no.~10, pp.~2000--2018, 2015.

\bibitem{friston2003dynamic}
K.~J. Friston, L.~Harrison, and W.~Penny, ``Dynamic causal modelling,'' {\em
  Neuroimage}, vol.~19, no.~4, pp.~1273--1302, 2003.

\bibitem{hochstein2002view}
S.~Hochstein and M.~Ahissar, ``View from the top: Hierarchies and reverse
  hierarchies in the visual system,'' {\em Neuron}, vol.~36, no.~5,
  pp.~791--804, 2002.

\bibitem{serre2007feedforward}
T.~Serre, A.~Oliva, and T.~Poggio, ``A feedforward architecture accounts for
  rapid categorization,'' {\em Proceedings of the national academy of
  sciences}, vol.~104, no.~15, pp.~6424--6429, 2007.

\bibitem{linde2019evidence}
J.~Linde-Domingo, M.~S. Treder, C.~Kerr{\'e}n, and M.~Wimber, ``Evidence that
  neural information flow is reversed between object perception and object
  reconstruction from memory,'' {\em Nature communications}, vol.~10, no.~1,
  pp.~1--13, 2019.

\bibitem{wheeler2013identification}
A.~L. Wheeler, C.~M. Teixeira, A.~H. Wang, X.~Xiong, N.~Kovacevic, J.~P. Lerch,
  A.~R. McIntosh, J.~Parkinson, and P.~W. Frankland, ``Identification of a
  functional connectome for long-term fear memory in mice,'' {\em PLoS Comput
  Biol}, vol.~9, no.~1, p.~e1002853, 2013.

\bibitem{squire1995retrograde}
L.~R. Squire and P.~Alvarez, ``Retrograde amnesia and memory consolidation: a
  neurobiological perspective,'' {\em Current opinion in neurobiology}, vol.~5,
  no.~2, pp.~169--177, 1995.

\bibitem{teyler2007hippocampal}
T.~J. Teyler and J.~W. Rudy, ``The hippocampal indexing theory and episodic
  memory: updating the index,'' {\em Hippocampus}, vol.~17, no.~12,
  pp.~1158--1169, 2007.

\bibitem{teyler1986hippocampal}
T.~J. Teyler and P.~DiScenna, ``The hippocampal memory indexing theory.,'' {\em
  Behavioral neuroscience}, vol.~100, no.~2, p.~147, 1986.

\bibitem{mcclelland1995there}
J.~L. McClelland, B.~L. McNaughton, and R.~C. O'Reilly, ``Why there are
  complementary learning systems in the hippocampus and neocortex: insights
  from the successes and failures of connectionist models of learning and
  memory.,'' {\em Psychological review}, vol.~102, no.~3, p.~419, 1995.

\bibitem{merkow2015human}
M.~B. Merkow, J.~F. Burke, and M.~J. Kahana, ``The human hippocampus
  contributes to both the recollection and familiarity components of
  recognition memory,'' {\em Proceedings of the National Academy of Sciences},
  vol.~112, no.~46, pp.~14378--14383, 2015.

\bibitem{danker2017trial}
J.~F. Danker, A.~Tompary, and L.~Davachi, ``Trial-by-trial hippocampal encoding
  activation predicts the fidelity of cortical reinstatement during subsequent
  retrieval,'' {\em Cerebral Cortex}, vol.~27, no.~7, pp.~3515--3524, 2017.

\bibitem{horner2015evidence}
A.~J. Horner, J.~A. Bisby, D.~Bush, W.-J. Lin, and N.~Burgess, ``Evidence for
  holistic episodic recollection via hippocampal pattern completion,'' {\em
  Nature communications}, vol.~6, no.~1, pp.~1--11, 2015.

\bibitem{staresina2013reversible}
B.~P. Staresina, E.~Cooper, and R.~N. Henson, ``Reversible information flow
  across the medial temporal lobe: the hippocampus links cortical modules
  during memory retrieval,'' {\em Journal of Neuroscience}, vol.~33, no.~35,
  pp.~14184--14192, 2013.

\bibitem{moscovitch2008hippocampus}
M.~Moscovitch, ``The hippocampus as a" stupid," domain-specific module:
  Implications for theories of recent and remote memory, and of imagination.,''
  {\em Canadian Journal of Experimental Psychology/Revue canadienne de
  psychologie exp{\'e}rimentale}, vol.~62, no.~1, p.~62, 2008.

\bibitem{georgopoulos1986neuronal}
A.~P. Georgopoulos, A.~B. Schwartz, and R.~E. Kettner, ``Neuronal population
  coding of movement direction,'' {\em Science}, vol.~233, no.~4771,
  pp.~1416--1419, 1986.

\bibitem{decharms2000neural}
R.~C. Decharms and A.~Zador, ``Neural representation and the cortical code,''
  {\em Annual review of neuroscience}, vol.~23, no.~1, pp.~613--647, 2000.

\bibitem{barlow1972single}
H.~B. Barlow, ``Single units and sensation: a neuron doctrine for perceptual
  psychology?,'' {\em Perception}, vol.~1, no.~4, pp.~371--394, 1972.

\bibitem{olshausen2004sparse}
B.~A. Olshausen and D.~J. Field, ``Sparse coding of sensory inputs,'' {\em
  Current opinion in neurobiology}, vol.~14, no.~4, pp.~481--487, 2004.

\bibitem{hubel1962receptive}
D.~H. Hubel and T.~N. Wiesel, ``Receptive fields, binocular interaction and
  functional architecture in the cat's visual cortex,'' {\em The Journal of
  physiology}, vol.~160, no.~1, p.~106, 1962.

\bibitem{mishkin1983object}
M.~Mishkin, L.~G. Ungerleider, and K.~A. Macko, ``Object vision and spatial
  vision: two cortical pathways,'' {\em Trends in neurosciences}, vol.~6,
  pp.~414--417, 1983.

\bibitem{gross1969visual}
C.~G. Gross, D.~B. Bender, and C.~d. Rocha-Miranda, ``Visual receptive fields
  of neurons in inferotemporal cortex of the monkey,'' {\em Science}, vol.~166,
  no.~3910, pp.~1303--1306, 1969.

\bibitem{tanaka1996inferotemporal}
K.~Tanaka, ``Inferotemporal cortex and object vision,'' {\em Annual review of
  neuroscience}, vol.~19, no.~1, pp.~109--139, 1996.

\bibitem{logothetis1996visual}
N.~K. Logothetis and D.~L. Sheinberg, ``Visual object recognition,'' {\em
  Annual review of neuroscience}, vol.~19, no.~1, pp.~577--621, 1996.

\bibitem{moser2008place}
E.~I. Moser, E.~Kropff, and M.-B. Moser, ``Place cells, grid cells, and the
  brain's spatial representation system,'' {\em Annu. Rev. Neurosci.}, vol.~31,
  pp.~69--89, 2008.

\bibitem{fried1997single}
I.~Fried, K.~A. MacDonald, and C.~L. Wilson, ``Single neuron activity in human
  hippocampus and amygdala during recognition of faces and objects,'' {\em
  Neuron}, vol.~18, no.~5, pp.~753--765, 1997.

\bibitem{kreiman2000category}
G.~Kreiman, C.~Koch, and I.~Fried, ``Category-specific visual responses of
  single neurons in the human medial temporal lobe,'' {\em Nature
  neuroscience}, vol.~3, no.~9, pp.~946--953, 2000.

\bibitem{quiroga2008sparsebutnot}
R.~Q. Quiroga, G.~Kreiman, C.~Koch, and I.~Fried, ``Sparse but not
  ‘grandmother-cell’coding in the medial temporal lobe,'' {\em Trends in
  cognitive sciences}, vol.~12, no.~3, pp.~87--91, 2008.

\bibitem{waydo2006sparse}
S.~Waydo, A.~Kraskov, R.~Q. Quiroga, I.~Fried, and C.~Koch, ``Sparse
  representation in the human medial temporal lobe,'' {\em Journal of
  Neuroscience}, vol.~26, no.~40, pp.~10232--10234, 2006.

\bibitem{quiroga2014single}
R.~Q. Quiroga, A.~Kraskov, F.~Mormann, I.~Fried, and C.~Koch, ``Single-cell
  responses to face adaptation in the human medial temporal lobe,'' {\em
  Neuron}, vol.~84, no.~2, pp.~363--369, 2014.

\bibitem{reddy2014concept}
L.~Reddy and S.~J. Thorpe, ``Concept cells through associative learning of
  high-level representations,'' {\em Neuron}, vol.~84, no.~2, pp.~248--251,
  2014.

\bibitem{christoff2016mind}
K.~Christoff, Z.~C. Irving, K.~C. Fox, R.~N. Spreng, and J.~R. Andrews-Hanna,
  ``Mind-wandering as spontaneous thought: a dynamic framework,'' {\em Nature
  Reviews Neuroscience}, vol.~17, no.~11, pp.~718--731, 2016.

\bibitem{kucyi2018just}
A.~Kucyi, ``Just a thought: How mind-wandering is represented in dynamic brain
  connectivity,'' {\em Neuroimage}, vol.~180, pp.~505--514, 2018.

\bibitem{addis2007remembering}
D.~R. Addis, A.~T. Wong, and D.~L. Schacter, ``Remembering the past and
  imagining the future: common and distinct neural substrates during event
  construction and elaboration,'' {\em Neuropsychologia}, vol.~45, no.~7,
  pp.~1363--1377, 2007.

\bibitem{tavanaei2018deep}
A.~Tavanaei, M.~Ghodrati, S.~R. Kheradpisheh, T.~Masquelier, and A.~S. Maida,
  ``Deep learning in spiking neural networks,'' {\em arXiv preprint
  arXiv:1804.08150}, 2018.

\bibitem{gerstner2002spiking}
W.~Gerstner and W.~M. Kistler, {\em Spiking neuron models: Single neurons,
  populations, plasticity}.
\newblock Cambridge university press, 2002.

\bibitem{ghosh2009spiking}
S.~Ghosh-Dastidar and H.~Adeli, ``Spiking neural networks,'' {\em International
  journal of neural systems}, vol.~19, no.~04, pp.~295--308, 2009.

\bibitem{taylor:stdp}
M.~Taylor, ``The problem of stimulus structure in the behavioural theory of
  perception,'' {\em South African Journal of Psychology}, vol.~3, pp.~23--45,
  1973.

\bibitem{caporale2008spike}
N.~Caporale and Y.~Dan, ``Spike timing--dependent plasticity: a hebbian
  learning rule,'' {\em Annu. Rev. Neurosci.}, vol.~31, pp.~25--46, 2008.

\bibitem{huang2014associative}
S.~Huang, C.~Rozas, M.~Trevino, J.~Contreras, S.~Yang, L.~Song, T.~Yoshioka,
  H.-K. Lee, and A.~Kirkwood, ``Associative hebbian synaptic plasticity in
  primate visual cortex,'' {\em Journal of Neuroscience}, vol.~34, no.~22,
  pp.~7575--7579, 2014.

\bibitem{mcmahon2012stimulus}
D.~B. McMahon and D.~A. Leopold, ``Stimulus timing-dependent plasticity in
  high-level vision,'' {\em Current biology}, vol.~22, no.~4, pp.~332--337,
  2012.

\bibitem{ott2020learning:sharing}
J.~Ott, E.~Linstead, N.~LaHaye, and P.~Baldi, ``Learning in the machine: To
  share or not to share?,'' {\em Neural Networks}, 2020.

\bibitem{masquelier2007unsupervised}
T.~Masquelier and S.~J. Thorpe, ``Unsupervised learning of visual features
  through spike timing dependent plasticity,'' {\em PLoS Comput Biol}, vol.~3,
  no.~2, p.~e31, 2007.

\bibitem{fei2007learning}
L.~Fei-Fei, R.~Fergus, and P.~Perona, ``Learning generative visual models from
  few training examples: An incremental bayesian approach tested on 101 object
  categories,'' {\em Computer vision and Image understanding}, vol.~106, no.~1,
  pp.~59--70, 2007.

\bibitem{berlucchi1999some}
G.~Berlucchi, ``Some aspects of the history of the law of dynamic polarization
  of the neuron. from william james to sherrington, from cajal and van
  gehuchten to golgi,'' {\em Journal of the History of the Neurosciences},
  vol.~8, no.~2, pp.~191--201, 1999.

\bibitem{barsalou2008grounded}
L.~W. Barsalou, ``Grounded cognition,'' {\em Annu. Rev. Psychol.}, vol.~59,
  pp.~617--645, 2008.

\bibitem{von1999and:binding}
C.~Von~der Malsburg, ``The what and why of binding: the modeler’s
  perspective,'' {\em Neuron}, vol.~24, no.~1, pp.~95--104, 1999.

\bibitem{treisman1980feature}
A.~M. Treisman and G.~Gelade, ``A feature-integration theory of attention,''
  {\em Cognitive psychology}, vol.~12, no.~1, pp.~97--136, 1980.

\bibitem{milner1974model}
P.~M. Milner, ``A model for visual shape recognition.,'' {\em Psychological
  review}, vol.~81, no.~6, p.~521, 1974.

\bibitem{shadlen1999synchrony}
M.~N. Shadlen and J.~A. Movshon, ``Synchrony unbound: a critical evaluation of
  the temporal binding hypothesis,'' {\em Neuron}, vol.~24, no.~1, pp.~67--77,
  1999.

\bibitem{Dalle:gpt3}
A.~Ramesh, M.~Pavlov, G.~Goh, and S.~Gray, ``Dall·e: Creating images from
  text.'' \url{https://openai.com/blog/dall-e/}.
\newblock Accessed: 2021-01-06.

\end{thebibliography}

\end{document}